\documentclass[12pt,preprint2]{aastex6}

\usepackage{natbib}
\usepackage{longtable}
\usepackage{enumerate}
\usepackage{amsmath}
\usepackage{hyperref}

\usepackage{CJK}

\def\msun{\ifmmode {\rm M}_{\mathord\odot}\else $M_{\mathord\odot}$\fi}
\def\rsun{\ifmmode {\rm R}_{\mathord\odot}\else $R_{\mathord\odot}$\fi}
\def\lsun{\ifmmode {\rm L}_{\mathord\odot}\else $L_{\mathord\odot}$\fi}
\newcommand{\hy}{\textsc{hyperion}}

\newcommand{\um}{$\mu$m}
\newcommand{\xmark}{\text{\sffamily X}}
\newcommand{\cmark}{\text{\sffamily \checkmark}}

\newcommand{\hii}{H\,{\sc ii}}
\newcommand{\Brut}{\textit{Brut}}

\slugcomment{}
\shorttitle{}
\shortauthors{}

\begin{document}
\begin{CJK*}{UTF8}{gbsn}    

\title{Assessing the Performance of a Machine Learning Algorithm in Identifying Bubbles in Dust Emission}

\author{Duo Xu(许铎)\altaffilmark{1}\altaffilmark{2}, Stella S. R. Offner\altaffilmark{1}\altaffilmark{2} }
\affil{
$^{1}$Department of Astronomy, University of Massachusetts, Amherst, MA 01002, USA;\\
$^{2}$Department of Astronomy, The University of Texas at Austin, Austin, TX 78712, USA;
\href{mailto:xuduo117@utexas.edu}{xuduo117@utexas.edu}, \href{mailto:soffner@astro.as.utexas.edu}{soffner@astro.as.utexas.edu}
}

\begin{abstract}
Stellar feedback created by radiation and winds from massive stars plays a significant role in both physical and chemical evolution of molecular clouds. This energy and momentum leaves an identifiable signature (``bubbles'') that affect the dynamics and structure of the cloud. Most bubble searches are performed ``by-eye", which are usually time-consuming, subjective and difficult to calibrate. Automatic classifications based on machine learning make it possible to perform systematic, quantifiable and repeatable searches for bubbles. We employ a previously developed machine learning algorithm, \Brut, and quantitatively evaluate its performance in identifying bubbles using synthetic dust observations. We adopt magneto-hydrodynamics simulations, which model stellar winds launching within turbulent molecular clouds, as an input to generate synthetic images. We use a publicly available three-dimensional dust continuum Monte-Carlo radiative transfer code, \hy, to generate synthetic images of bubbles in three Spitzer bands (4.5 \um, 8 \um\ and 24 \um). We designate half of our synthetic bubbles as a training set, which we use to train \Brut\ along with citizen-science data from the Milky Way Project. We then assess \Brut's accuracy using the remaining synthetic observations. We find that after retraining \Brut's performance increases significantly, and it is able to identify yellow bubbles, which are likely associated with B-type stars. \Brut\ continues to perform well on previously identified high-score bubbles, and over 10\% of the Milky Way Project bubbles are reclassified as high-confidence bubbles, which were previously marginal or ambiguous detections in the Milky Way Project data. We also investigate the size of the training set, dust model, evolution stage and background noise on bubble identification. 

\end{abstract}


\section{Introduction}

During the process of star formation, stellar feedback plays a significant role in both physical and chemical evolution of molecular clouds \citep{1999RvMP...71..173H, 2014prpl.conf..451F}. One of the most important feedback mechanisms is mass-loss \citep{1985ARA&A..23..267L}. There are two typical manifestations of stellar winds: protostellar outflows, which are often highly collimated, and radiatively driven winds from main sequence stars, which are more isotropic \citep{2006ApJ...649..759C, 2010ApJ...715.1170A, 2011ApJ...742..105A, 2015ApJS..219...20L}. Both type of stellar winds inject momentum and energy into the environment, and thereby affect the dynamics and structure of the parent molecular cloud. 

Recent observational studies have shown that the momentum and energy injected by stellar winds are one or more orders of magnitude larger than those of outflows owing to their larger volume and longer lifetime \citep{2011ApJ...742..105A, 2015ApJS..219...20L}. \citet{2011ApJ...742..105A} found that the energy injection rate from these stellar winds is comparable to the turbulent dissipation rate in the Perseus molecular cloud, which means that in the current epoch, stellar feedback is sufficient to maintain the observed turbulence in Perseus. A similar conclusion was also reached by \citet{2015ApJS..219...20L} in the Taurus molecular cloud. It is notable that both regions are low-mass star forming regions, and high-mass stars, which generally dominate feedback energetics are absent.

Simulations confirm the significant kinematic impact due to stellar feedback on the global star formation process. Winds can replenish energy dissipated by turbulence and also trigger star formation by compressing the cloud \citep{2008MNRAS.391....2D,2002ApJ...566..302M,2005MNRAS.358..291D,2013MNRAS.436.3430D,2014MNRAS.442..694D,2007ApJ...662..395N,2010ApJ...709...27W}. Winds can also gradually ablate the molecular material from forming stellar clusters \citep{2013MNRAS.431.1337R}. \citet{2015ApJ...811..146O} quantified the stellar wind mass-loss rates for individual stars, which they found must be greater than $10^{-7}$ \msun\, yr$^{-1}$ to be consistent with observations. Additionally, ionizing radiation feedback from O-stars also influences the morphology of clouds and the formation of stars \citep{2005MNRAS.358..291D,2013MNRAS.436.3430D,2014MNRAS.442..694D,2015MNRAS.448.3248G,2016ApJ...819..137K}.

Despite many observational and theoretical studies, the importance and impact of feedback on molecular clouds remain debated. This is because wind signatures are difficult to identify and quantify. Most bubble searches are done ``by eye'' \citep{2006ApJ...649..759C, 2011ApJ...742..105A, 2015ApJS..219...20L}. For example, over 35,000 citizen scientists participated in the Milky Way Project \citep[MWP,][]{2012MNRAS.424.2442S} in order to identify bubbles in Spitzer images. This approach is time-consuming, subjective and difficult to calibrate \citep{2014ApJS..214....3B}. Analyzing the completeness of visually identified bubbles, which has a significant effect on the estimation of the injected momentum and energy, remains a great challenge. However, automatic classifications driven by machine learning approaches enable systematic, quantifiable and repeatable searches to identify bubbles \citep{2011ApJ...741...14B, 2014ApJS..214....3B}.

One of the most popular types of machine learning algorithms in astronomical classification is ``Random Forests'' \citep[e.g.][]{2010ApJ...712..511C, 2014ApJS..214....3B, 2014AJ....148...21M}, which are based on decision trees. A decision tree is a data structure which classifies feature vectors by computing a series of constraints, and propagating vectors down the tree based on whether these constraints are satisfied. Compared to other machine learning approaches, the Random Forests approach does well in classifying problems that have a large number of feature dimensions \citep{Breiman2001}. \citet{2014ApJS..214....3B} developed an algorithm \Brut\ based on Random Forests and applied it to classifying bubbles in the Milky Way. For each bubble, they defined a ``score'', which is related to the probability that a given structure is a bubble. After conducting a blind search in the Milky Way, they found a substantial population of low-score bubble candidates  not in MWP catalog produced by citizen scientists. In other words, citizen scientists are likely to miss a significant number of bubbles, but machine leaning can compensate for some of this incompleteness. 

Increasingly rich and detailed data of the local ISM and star-forming regions are available, such as GLIMPSE \citep[Galactic Legacy Infrared Mid-Plane Survey Extraordinaire,][]{2003PASP..115..953B}, Hi-GAL \citep[Herschel infrared Galactic Plane,][]{2010PASP..122..314M} Survey and GALFA-HI \citep[The Galactic Arecibo L-band Feed Array HI,][]{2011ApJS..194...20P} Survey. Parsing these extensive data visually is prohibitively time-consuming but is possible with the aid of machine learning algorithms.

There are two main types of machine learning algorithms: unsupervised learning and supervised learning. Unsupervised learning algorithms make their own criteria to discover structure in the data.
An algorithm that learns from a training dataset and makes decisions based on the input ``knowledge'' is called supervised learning. Supervised learning iteratively makes predictions on the training data and is corrected by the input training dataset. Consequently, the training dataset plays a significant role in the performance accuracy. 

One fundamental problem with visual identification is that bubbles identified ``by eye" are not objective and can be incorrect, such that machine learning approaches trained using flawed visual data will in turn produce defective identifications. In addition, there is no independent, quantitative assessment for completeness or any clear metric to determine how well bubbles are actually identified. One solution is to use realistic simulations, where feedback properties are known and well-defined. Such simulations can evaluate the accuracy of the training data and, in turn, supplement the original training dataset.

In this paper, we assess the performance of \Brut\ in identifying bubbles using synthetic observations. We produce synthetic dust observations of bubbles in simulations. We use these as a supplemental training set to retrain \Brut\ and test the performance of retrained \Brut\ in classifying both synthetic bubbles and observed bubbles. We describe the method we use to construct synthetic observations and the details of the machine learning algorithm in Section~\ref{Methods}. We compare and discuss several synthetic observation models in Section~\ref{Synthetic Observations}. In Section~\ref{Result}, we present the performance of retrained \Brut\ in classifying both synthetic bubbles and observed bubbles. We summarize our results and conclusions in Section~\ref{Conclusion}.

\section{Methods}
\label{Methods}

\subsection{Hydrodynamic Simulations}

\label{Hydrodynamic Simulations}
We adopt the magneto-hydrodynamics simulations from \citet{2015ApJ...811..146O}, which aim to model winds from intermediate-mass stars and explore their impact on cloud morphology and turbulence. The simulations model a piece of a molecular cloud with length of $L=5$ pc, mass of $M=3762~\msun$ and periodic boundary conditions. The initial cloud temperature is $T= 10$ K. {The initial density and velocity conditions are set through driving the gas without gravity by adding random large-scale perturbations to the velocity field. These simulations share the same Alfv\'en Mach number 2.3 but their magnetic field distributions 
are spatially different at the initial time. Their velocity and density Fourier spectral slopes are comparable to $S(k)\propto k^{-1.7}$ and $S(k)\propto k^{-1.3}$, respectively. 
The turbulence is initially external driving but ceases when the stellar sources are inserted and the begin feedbacks.} Table~\ref{Model Properties} lists the parameters of these models. More details about the simulations can be found in \citet{2015ApJ...811..146O}.

We adopt outputs from the strong wind run in which five stellar sources with different mass-loss rates are randomly placed. The number density of sources is similar to that in Perseus. These sources are all B-type stars with the mass-loss rates ranging from 2.6$\times10^{-8}$ -- 1.8$\times10^{-5}$ \msun\ yr$^{-1}$. Table~\ref{stellar-sources} lists the physical parameters of each of the five stellar sources. In this work, we explored outputs with different evolution stages and different turbulence realizations.

\begin{table}[]
\begin{center}
\caption{Physical Parameters of the Stellar Sources   \label{stellar-sources}}
\begin{tabular}{ccccc}
\hline
ID & $M$ (\msun) & $L$ ($10^{3}$ \lsun) & $T$ ($10^{4}$ K) & $\dot{M}$ ($10^{-7} $\msun yr$^{-1}$) \\
 \hline
    1 & 3.8     & 0.19  & 2.3   & 0.35 \\
    2 & 10.4   & 6.3   & 3.8   & 9.1 \\
    3 & 12.2    & 10.3  & 3.6   & 17.7 \\
    4 & 13.1   & 12.8  & 3.1   & 12.4 \\
    5 & 12.4   & 10.8  & 2.6   & 2.5 \\
\hline

\end{tabular}
\end{center}
\end{table}

\begin{table}[]
\begin{center}
\caption{Model Properties$^{a}$ \label{Model Properties}}
\begin{tabular}{ccc}
\hline
Model & $t_{\rm i}$ ($t_{\rm{cross}}$) & $t_{\rm run}$ (Myr)   \\
 \hline
    T1\_t1$^{b}$    & 1.6   & 0.1   \\
    T2\_t1$^{c}$ & 2.0  & 0.1   \\
    T2\_t0   & 2.0  & 0.05   \\
\hline
\multicolumn{3}{p{0.65\linewidth}}{Notes:}\\
\multicolumn{3}{p{0.65\linewidth}}{
$^{a}$ Model name, the initial start time in crossing times and the evolutionary time. All models have $L$ = 5 pc, $M$ = 3762 \msun\, $T_{i}$ = 10 K and initial $B$=13.5 $\mu G$.}\\
\multicolumn{3}{p{0.65\linewidth}}{
$^{b}$ Output corresponding to the model ``W1\_T1'' with an evolutionary time of 0.1 Myr in \citet{2015ApJ...811..146O}.}\\ 
\multicolumn{3}{p{0.65\linewidth}}{
$^{c}$ Output corresponding to the model ``W1\_T2'' in \citet{2015ApJ...811..146O}.}
\end{tabular}
\end{center}
\end{table}

\subsection{Hyperion} 
\label{Hyperion}

We use the publicly available three-dimensional dust continuum Monte-Carlo radiative transfer code \hy\ \citep{2011A&A...536A..79R} to generate synthetic observations of the simulations described in Section~\ref{Hydrodynamic Simulations}. We adopt the gas density and temperature distributions from the outputs listed in Table~\ref{Model Properties} and the stellar properties from Table~\ref{stellar-sources} as inputs. \hy\ assumes stars radiate as a blackbody.

\begin{figure*}[htp]
\centering
\includegraphics[width=1.0\linewidth]{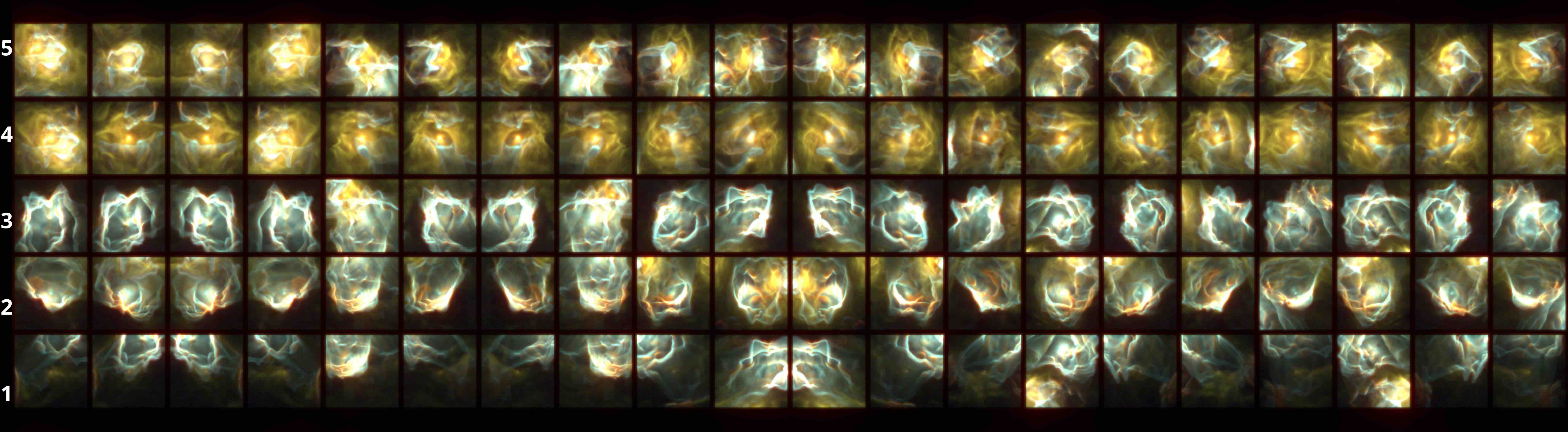}
\caption{Three-color synthetic images of five sources with 20 different viewing angles adopting the dust model in \citet{2016arXiv160302270K}. Red, green and blue represents 24 \um, 8 \um\ and 4.5 \um\ emission, respectively. }
\label{fig.Koepferlbubble}
\end{figure*}

Assumptions about the dust properties strongly influence the resulting emission. A variety of models for ISM dust have been proposed in the literature \citep[e.g.][]{1994ApJ...422..164K,2003ARA&A..41..241D,2016arXiv160302270K}, and we explore four different models in this work. Following \citet{2016arXiv160302270K}, we combine three different dust grain models with 80.63\% big grains ($>$200 \AA), 13.51\% smaller dust species, called very small grains (20--200 \AA, vsg), and 5.86\% PAH molecules, called ultra-small grains ($<$20 \AA, usg). We label this dust model  ``K16'' in the following discussion. We assume a moderate gas-to-dust ratio of 100 \citep{1979ARA&A..17...73S} and adopt a regular Cartesian grid with young stars embedded within. We calculate the emission for 20 different angular views and convolve the spectra with the Spitzer transmission curve \citep{2004SPIE.5487..244Q, MIPS...Instrument} to generate synthetic images in three Spitzer bands (4.5, 8, 24 \um). Figure~\ref{fig.Koepferlbubble} shows synthetic bubble images of the five sources with 20 different viewing angles.

In addition to the K16 dust model above, we adopt three other commonly used dust models to produce synthetic observations:
\begin{enumerate}[(1).]
\item ``kmh'' dust model \citep{1994ApJ...422..164K}, which consists of astronomical silicates, graphite, and carbon with full scattering properties,

\item ``Draine'' dust model \citep{2003ARA&A..41..241D}, which is mainly Milky-Way carbonaceous-silicate grains,

\item ``IPS'' dust model \citep{2003A&A...410..611S}, which represents ``iron-poor'' silicate dust.
\end{enumerate}

\begin{figure*}[htp]
\centering
\includegraphics[width=1.0\linewidth]{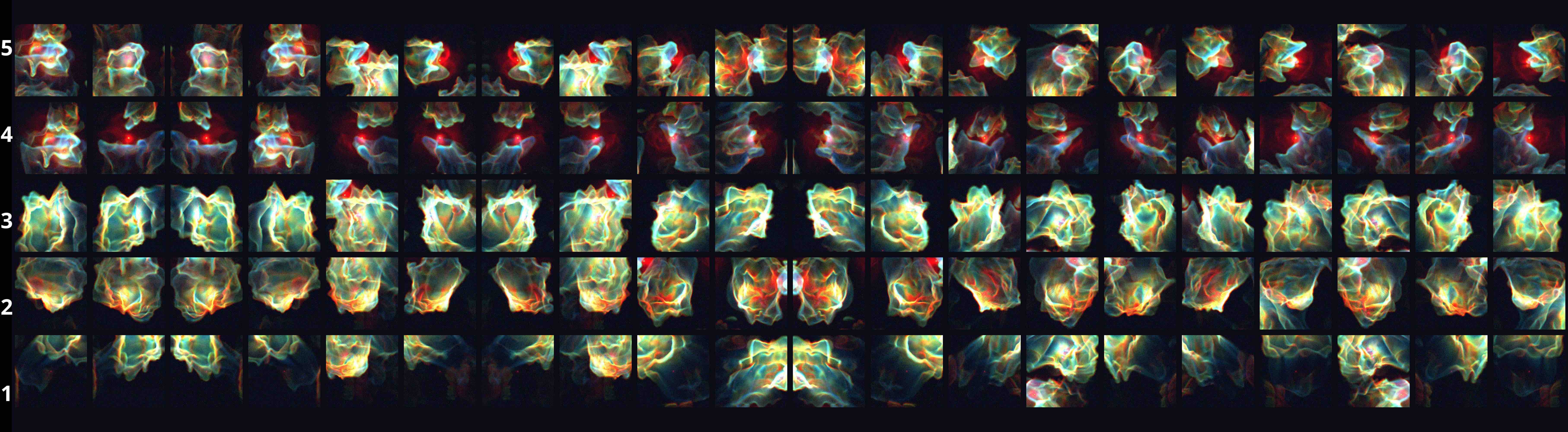}
\caption{Same as Figure~\ref{fig.Koepferlbubble} but adopting the kmh dust model in \citet{1994ApJ...422..164K}.} 
\label{fig.KMHbubble}
\end{figure*}

Figure~\ref{fig.KMHbubble} shows synthetic images adopting the kmh dust model. The synthetic observations adopting the Draine and IPS dust models are similar to those adopting the kmh dust model, so we only include images with the kmh model. 

\begin{figure}[htp]
\centering
\includegraphics[width=1.0\linewidth]{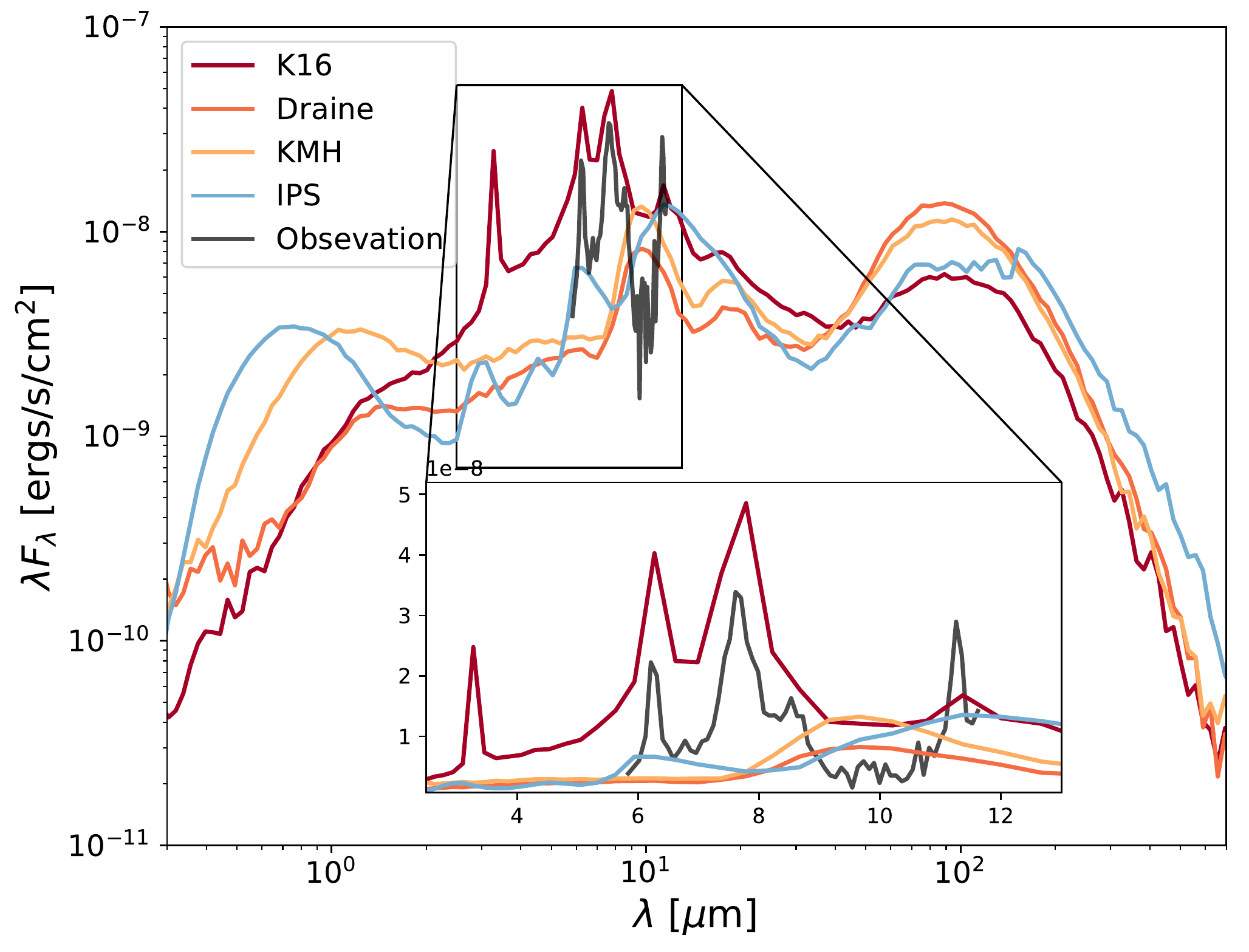}
\caption{The SEDs of different dust models compared with the spectra of the Ophiuchus cloud LDN 1688 observed by \citet{2013MNRAS.428.2617R}. }
\label{fig.allsed}
\end{figure}

The SEDs of different dust models show distinct differences, especially at 8 \um\  where PAH emission dominates. We extract the observed spectra of the main molecular cloud of Ophiuchus, LDN 1688 \citep{2013MNRAS.428.2617R} and compare the SEDs of the different dust models as shown in Figure~\ref{fig.allsed}. The K16 dust model appears to be more realistic since it includes PAH emission while the other models lack PAH emission around 8 \um. Since the SEDs of the kmh model, Draine model, and the IPS model have a similar intensity at 4.5 \um, 8 \um\ and 24 \um, the Draine and IPS three-color synthetic images look similar to the kmh model shown in Figure~\ref{fig.KMHbubble}. The interiors of the bubbles in Figure~\ref{fig.KMHbubble} appear to be redder. This is because the 24 \um\ emission is stronger, but they lack 8 \um\ emission, compared to Figure~\ref{fig.Koepferlbubble}. Consequently, we adopt the K16 dust model for the remainder of the analysis.

\subsection{\Brut}
\label{Brut}
\Brut\ is a machine learning algorithm developed to identify bubbles in infrared images of the Galactic midplane \citep{2014ApJS..214....3B}. \Brut\ uses a Random Forest approach that is based on decision trees. A decision tree is a data structure that classifies a set of features, i.e., a numerical vector that describes the properties of each region. \Brut\ computes a series of constraints and propagates the features down the tree based on whether these constraints are satisfied.  \Brut\ defines four features, which extract the most useful information about the difference between bubble and non-bubble images. It concatenates them into a single feature vector to carry out the classification. 

\citet{2014ApJS..214....3B} adopted bubbles identified by citizen scientists from the Milky Way Project as a training set. We include this same data for our analysis. The training set consists of 468 visually identified bubbles and 2289 random fields that are not centered on a bubble. \Brut\ has three forests on different subsets of the sky, which we denote r1, r2 and r3. Each forest is trained using examples from two-thirds of the survey area and then tested using the remaining one-third area, as shown in Table~\ref{Random Forest Zone}. The illustration of the zones can be found in Figure 7 in  \citet{2014ApJS..214....3B}.

After training, \Brut\ returns a score related to the probability that a given structure is a bubble. If $P$ is the probability that a given structure belongs to the bubble set, the \Brut\ score is defined as $2P-1$, where -1 is unlikely to be a bubble and +1 is very likely. To find the threshold score for true bubbles, \citet{2014ApJS..214....3B} conduct a survey using experienced astronomers. They find about 50\% of astronomers are likely to judge a region with a \Brut\ score of 0.2 as a bubble. Consequently, they set 0.2 as the minimum acceptable score.

\begin{table*}[]
\begin{center}
\caption{Random Forest Zone \label{Random Forest Zone}}
\begin{tabular}{ccc}
\hline
Random Forest Name & Training Zone ($l$) $^{a}$ &  Test Zone ($l$)  \\
 \hline
    r1    &  $3n+0.5^{\circ} \le l < 3n+1.5^{\circ}  $ $^{b}$  & $3n+1.5^{\circ} \le l < 3n+3.5^{\circ} $ $^{d}$  \\
    r2   &  $3n+1.5^{\circ} \le l < 3n+2.5^{\circ} $  & $3n-0.5^{\circ} \le l < 3n+1.5^{\circ} $   \\
    r3   &  $3n-0.5^{\circ} \le l < 3n+0.5^{\circ} $ $^{c}$ & $3n+0.5^{\circ} \le l < 3n+2.5^{\circ} $  \\
\hline
\multicolumn{3}{p{0.65\linewidth}}{Notes:}\\
\multicolumn{3}{p{0.65\linewidth}}{
$^{a}$  The training zones are interleaved across all longitudes.}\\
\multicolumn{3}{p{0.65\linewidth}}{
$^{b}$  $n$ is an integer ranging from 0 to 119.}\\
\multicolumn{3}{p{0.65\linewidth}}{
$^{c}$  When $n$ is 0, the training zone is $359.5^{\circ} (-0.5^{\circ}) \le l < 0.5^{\circ} $}\\
\multicolumn{3}{p{0.65\linewidth}}{
$^{d}$  When $n$ is 119, the test zone is $358.5^{\circ} \le l < 0.5^{\circ} (360.5^{\circ}) $}\\

\end{tabular}

\end{center}
\end{table*}

\section{Synthetic Observations}
\label{Synthetic Observations}

We adopt models with different evolutionary stages and turbulence properties as listed in Table~\ref{Model Properties} and consider different dust models in the synthetic observations.

\subsection{Cropped Data}
\label{Cropped Data}

When carrying out the synthetic observations, we exploit the periodic nature of the simulation domain and wrap the data so all views have complete $N^3$ voxels, {where $N$ is the number of pixels in one dimension}. However, for large image sizes ($L\ge 3$ pc), the Monte Carlo calculation becomes prohibitively expensive at the resolution we require. Instead, we crop the data into cubes of length 2.2 pc and 3 pc with each individual stellar object at the center.

\begin{figure*}[htp]
\centering
\includegraphics[width=1.0\linewidth]{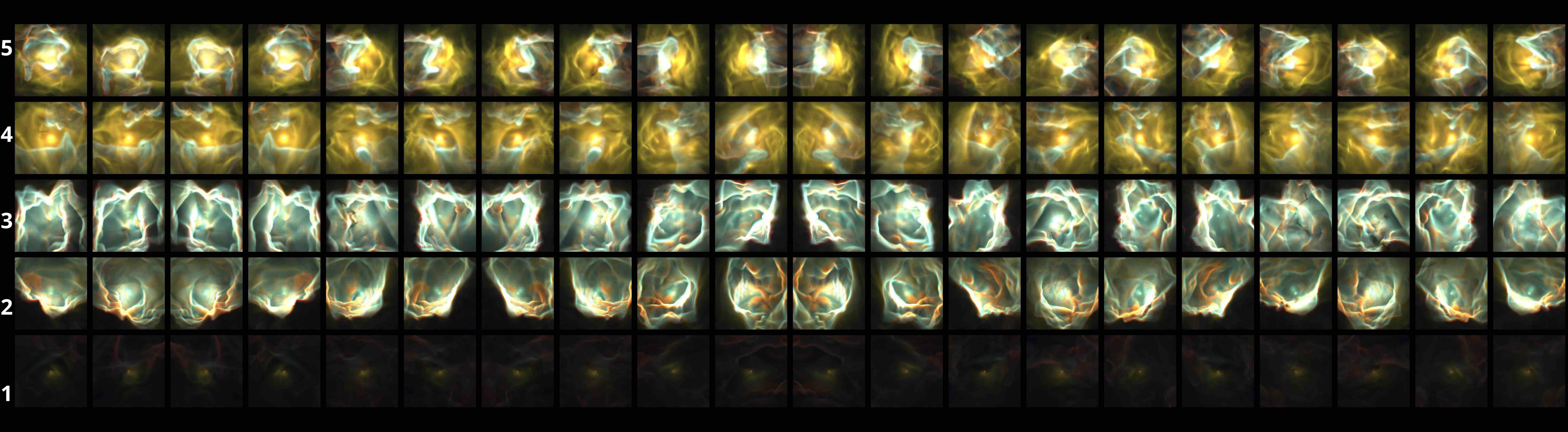}
\caption{Three-color synthetic images adopting the K16 dust model where the \hy\ input is cropped to 2.2 pc. }
\label{fig.bubble1.1}
\end{figure*}

\begin{figure*}[htp]
\centering
\includegraphics[width=1.0\linewidth]{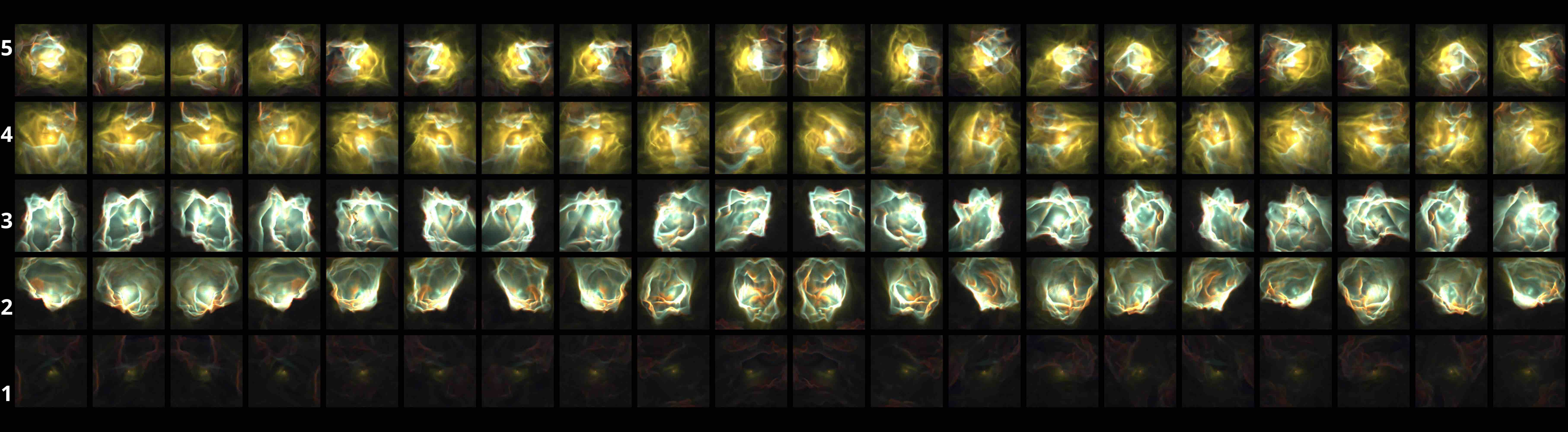}
\caption{Three-color synthetic images adopting the K16 dust model where the \hy\ input is cropped to 3 pc. }
\label{fig.bubble1.5}
\end{figure*}

Figure~\ref{fig.bubble1.1} and \ref{fig.bubble1.5} show the cropped synthetic bubble images. Compared with Figure~\ref{fig.Koepferlbubble}, in which the bubbles are embedded in the molecular cloud, the synthetic bubble images of the cropped data (Figure~\ref{fig.bubble1.1} and \ref{fig.bubble1.5}) are less extincted but the morphology does not change significantly. They appear to be a little bit brighter and bluer, which means the shorter wavelength emission is less attenuated. Another advantage of this strategy is that the synthetic bubble images are not contaminated by as much foreground or background emission. For example, the bottom row of Figure~\ref{fig.Koepferlbubble} is contaminated by the bubble from the source in the third row. Although observational data likely have overlapping bubbles, most bubbles identified by citizen scientists in MWP tend to be isolated. Consequently, we adopt the cropping strategy to generate the synthetic bubble images in the following discussion.

\subsection{Evolutionary Stage}
\label{Evolutionary Stage}

\begin{figure*}[htp]
\centering
\includegraphics[width=1.0\linewidth]{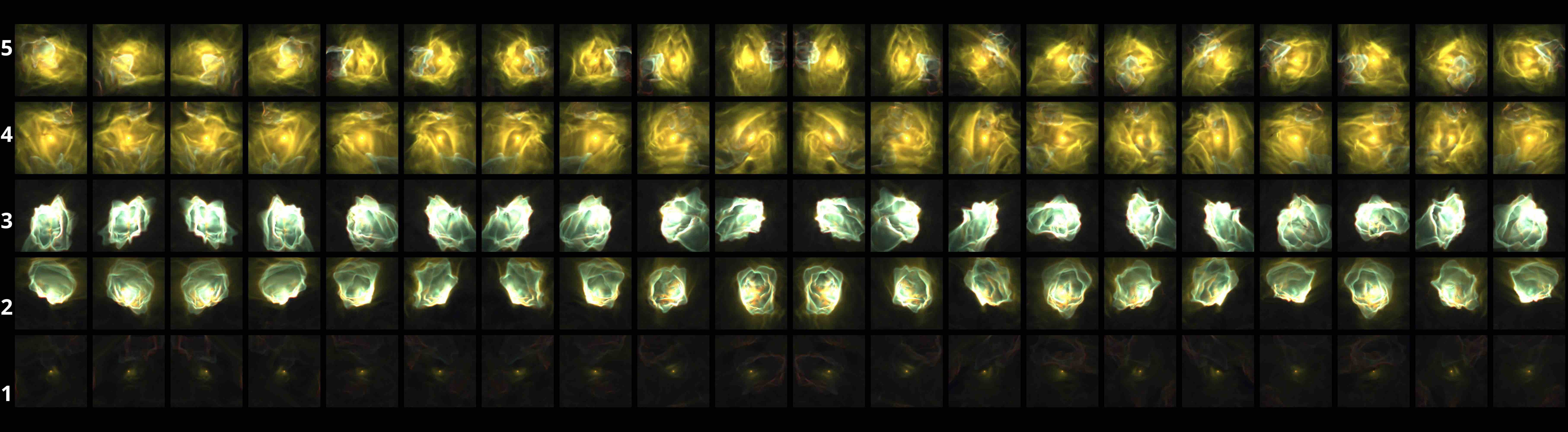}
\caption{Three-color synthetic images at an earlier evolution stage with 0.05 Myr (``T2\_t0'' listed in Table~\ref{Model Properties}) where the \hy\ input is cropped to 3 pc. } 
\label{fig.bubble1.5-2155}
\end{figure*}

The morphology of the bubbles changes with time as the winds expand into the cloud and interact with the surrounding gas. At earlier evolutionary stages, the bubbles are more compact compared with those at later stages, which have undergone additional expansion driven by the stellar wind. Figure~\ref{fig.bubble1.5-2155} shows younger bubbles (``T2\_t0'' listed in Table~\ref{Model Properties}). The bubbles at the earlier time appear brighter in the center, owing to their compact and concentrated structure. 

\subsection{Turbulent Realization}
\label{Turbulent Realization}

\begin{figure*}[htp]
\centering
\includegraphics[width=1.0\linewidth]{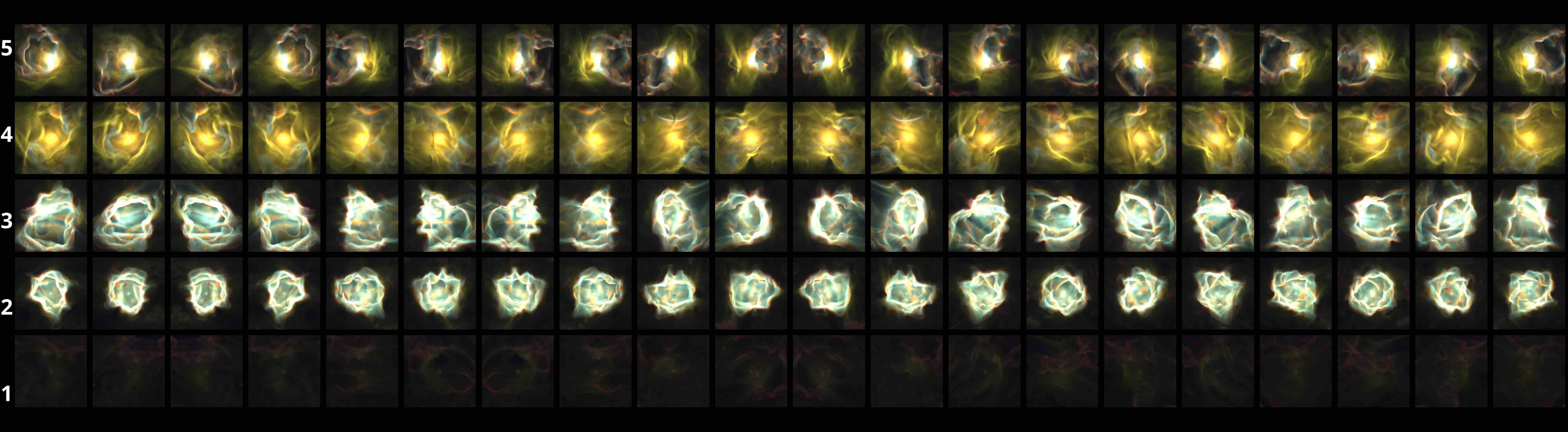}
\caption{Three-color synthetic images with different turbulence where the \hy\ input is cropped to 3 pc.}
\label{fig.bubble1.5-data}
\end{figure*}

We also analyze a simulation with different initial turbulence. The synthetic observation process remains the same as described above, where we crop the \hy\ input data cube and use the K16 dust model. Figure~\ref{fig.bubble1.5-data} shows the synthetic images with different initial turbulence (``T1\_t1'' listed in Table~\ref{Model Properties}). 

``T1\_t1'' and ``T2\_t1'' have the same initial mean magnetic field, ratio of thermal to magnetic pressure, mean density, and stellar properties, but the shape of the bubbles are distinctly different owing to the different density distribution of the cloud material. Since the turbulent structure of real molecular clouds is varied, we adopt different initial turbulence to explore the diversity of bubble morphology and enrich our training dataset.

\subsection{Noise}
\label{Noise}

\begin{figure*}[htp]
\centering
\includegraphics[width=1.0\linewidth]{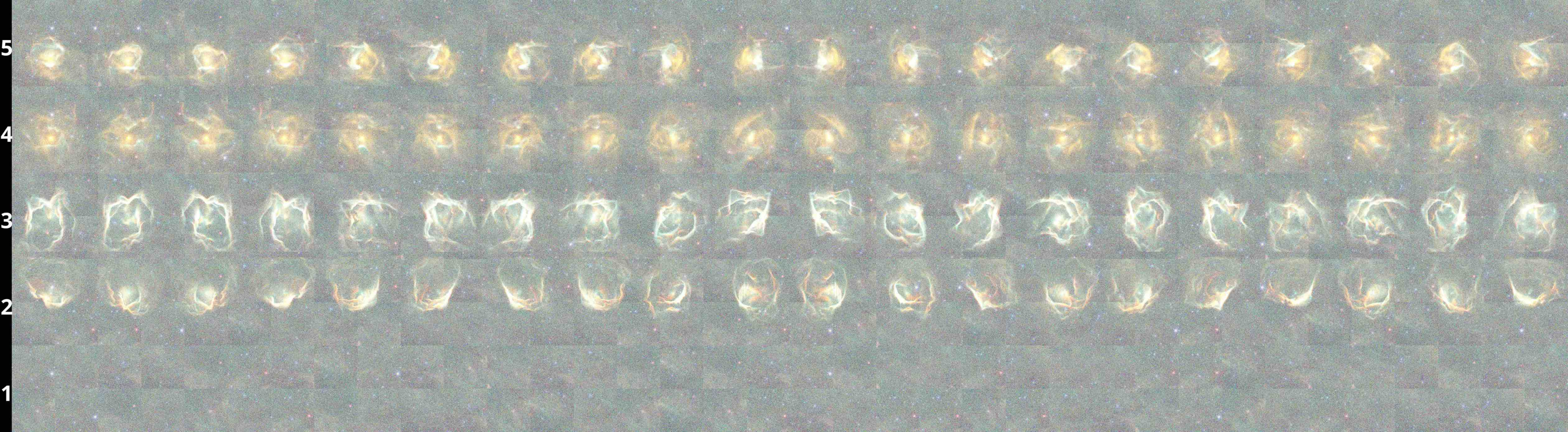}
\caption{Three-color synthetic images with noise. From top to bottom: Figure~\ref{fig.Koepferlbubble}, \ref{fig.bubble1.1} and \ref{fig.bubble1.5} with noise added.}
\label{fig.bubblenoise1}
\end{figure*}

The synthetic images are smooth, which is distinct from real observational images, which have fluctuations produced by noise. It is important that the training data be as close as possible to the observational data to reduce bias in detection caused by differences. To make the synthetic images more realistic, we identify patches of {GLIMPSE} data that are removed from the Galactic plane and have low signal to noise (S/N). We add these ``stamps” to the synthetic images using the same S/N as the {GLIMPSE} data, where S/N $\sim 8$. Figure~\ref{fig.bubblenoise1} shows the synthetic bubble images with noise.

\section{Results}
\label{Result}

\subsection{Retraining \Brut\ with Synthetic Observation}
\label{Retrain Brut}

We divide all the synthetic images into two equal parts. One half acts as a training data set, which we use to supplement the original MWP bubble set. The remainder serve as a test set, which allows us to assess the performance of the retrained algorithm. We summarize all the synthetic images we use in the training and testing sets in Table~\ref{bubble-parameter}.

\begin{figure}[htp]
\centering
\includegraphics[width=.90\linewidth]{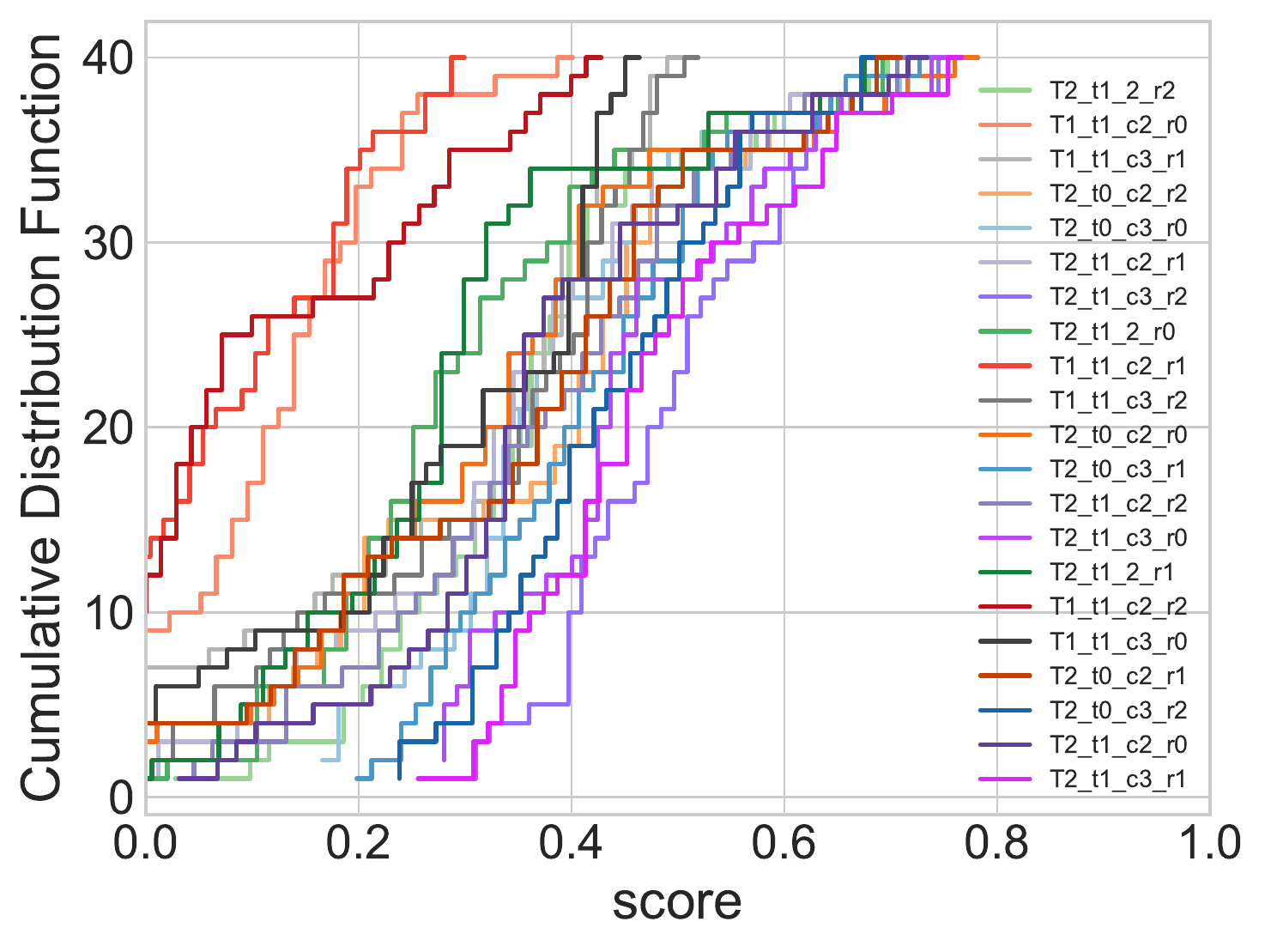}
\includegraphics[width=.90\linewidth]{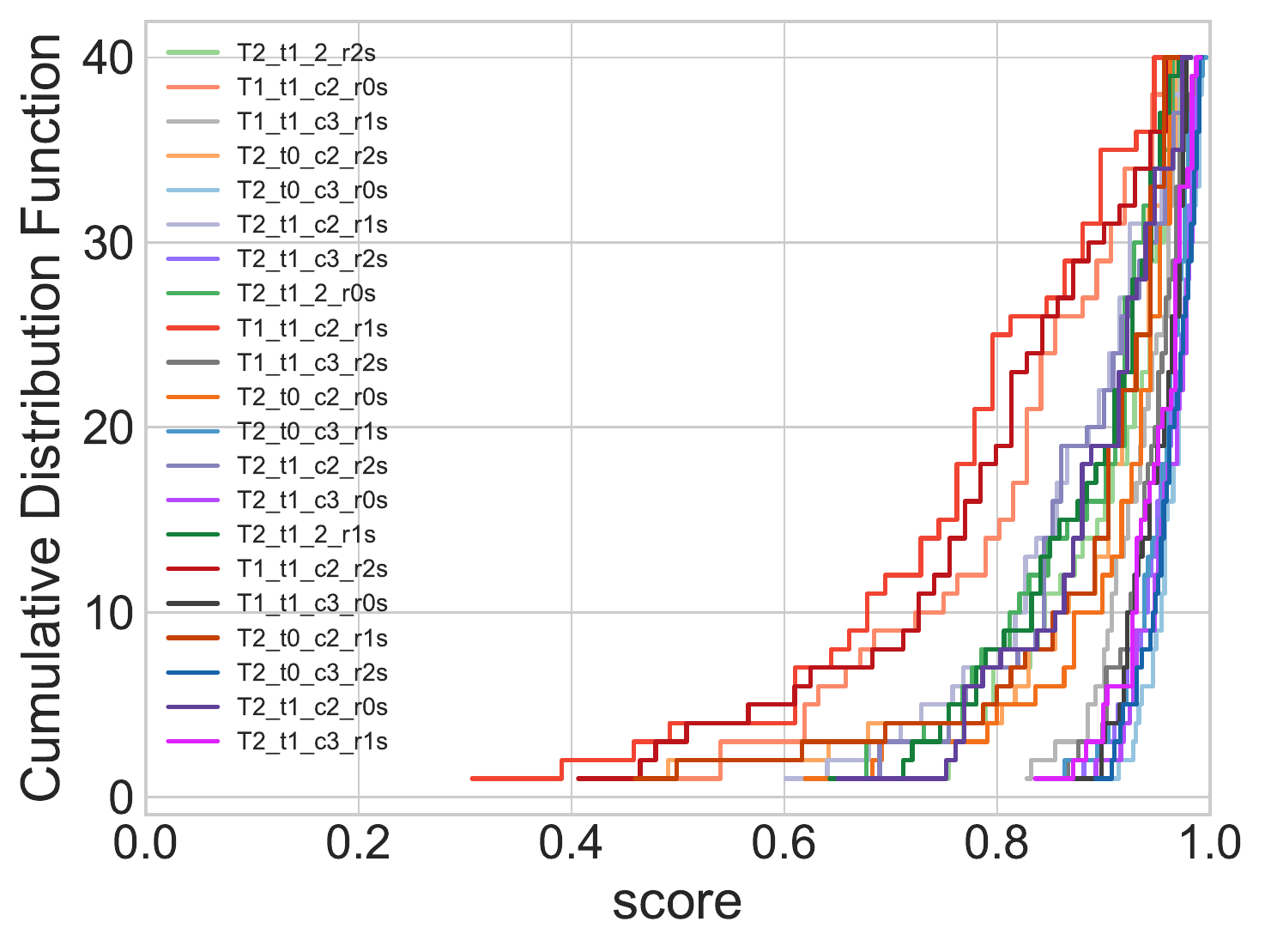}
\caption{The cumulative distribution function (CDF) of all the scores given by Brut with the original {training}  (top panel) and the algorithm retrained on noiseless synthetic images (bottom panel). The labels are described in Table~\ref{parameters of rf}. }
\label{fig.socre-cdf-nonoise}
\end{figure}

We analyze the performance of the three Random Forests before and after supplementing with the new training data. First, we retrain \Brut\ using the synthetic images without noise (IDs 1-7 in Table~\ref{bubble-parameter}).   
Figure~\ref{fig.socre-cdf-nonoise} shows the performance with the original {training} and the algorithm retrained on noiseless synthetic images on the test bubbles. Table~\ref{parameters of rf} briefly describes the meaning of labels in Figure~\ref{fig.socre-cdf-nonoise}. The scores returned after retraining on the noiseless data are significantly higher than those given by the original {training}. After {retraining}, the feature vector more accurately represents the synthetic bubbles and \Brut\ does a better job identifying them.

\begin{figure}[htp]
\centering
\includegraphics[width=.90\linewidth]{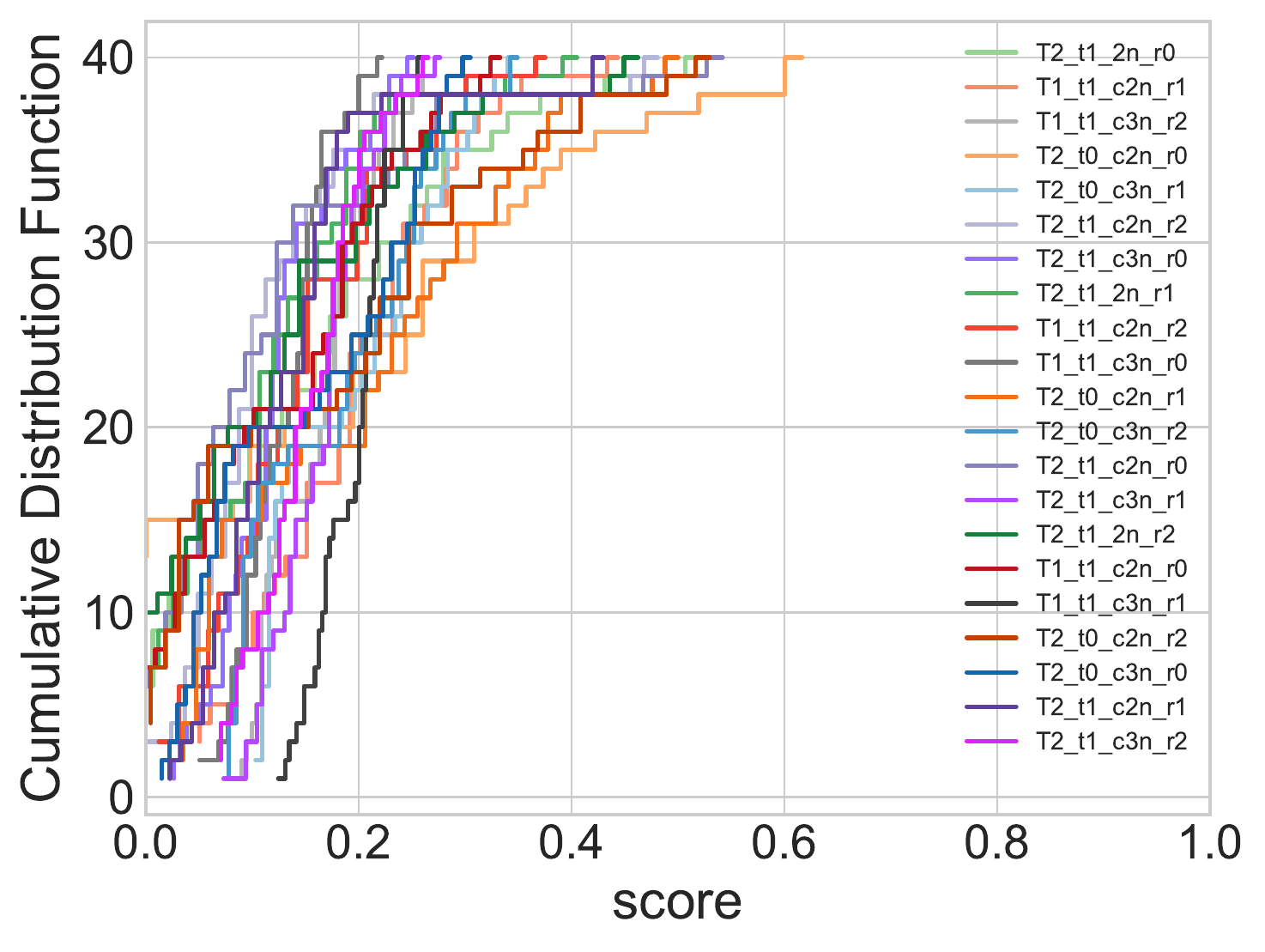}
\includegraphics[width=.90\linewidth]{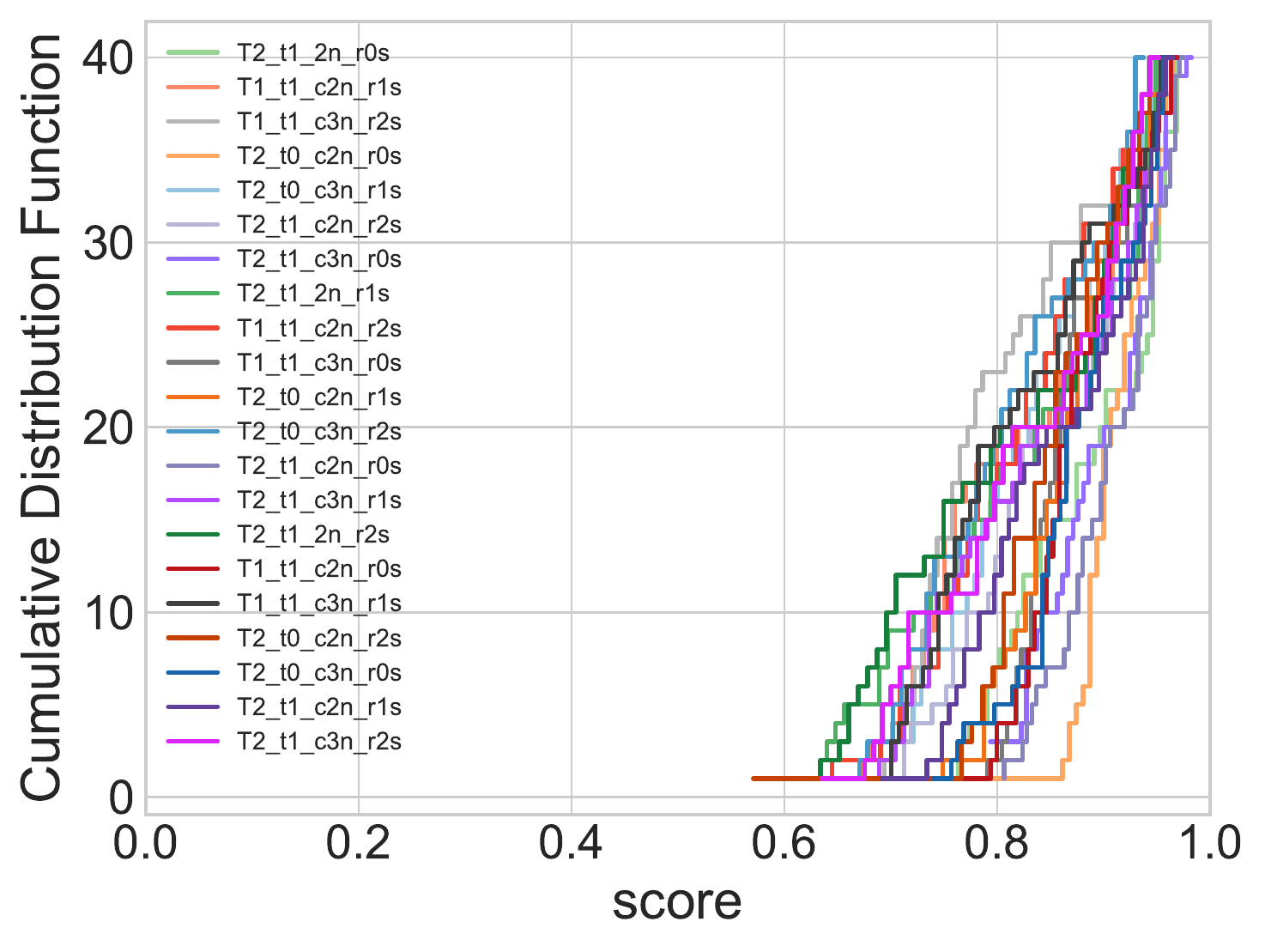}
\caption{The CDF of all the scores given by Brut with the original {training}  (top panel) and the algorithm retrained on synthetic images with and without noise (bottom panel). The labels are described in Table~\ref{parameters of rf}. }
\label{fig.socre-cdf-noise}
\end{figure}

We then augment the training set by adding the bubbles with IDs 7-14 in Table~\ref{bubble-parameter}, so that the new training set consists of half of the bubbles with and without noise. Figure~\ref{fig.socre-cdf-noise} shows the performance with the original {training} and the algorithm retrained on synthetic images with and without noise on the second half of the noisy data. The scores returned by the retrained algorithm are significantly higher than those given with the original {training}. Compared with the scores retrained using noiseless data in Figure~\ref{fig.socre-cdf-nonoise}, the scores given by the retrained algorithm including some noisy images are more concentrated. This is likely because delicate bubble structure is reduced, i.e., there is less variation in bubble appearance since the noise hides small-scale sub-structure. 

\begin{figure}[htp]
\centering
\includegraphics[width=.90\linewidth]{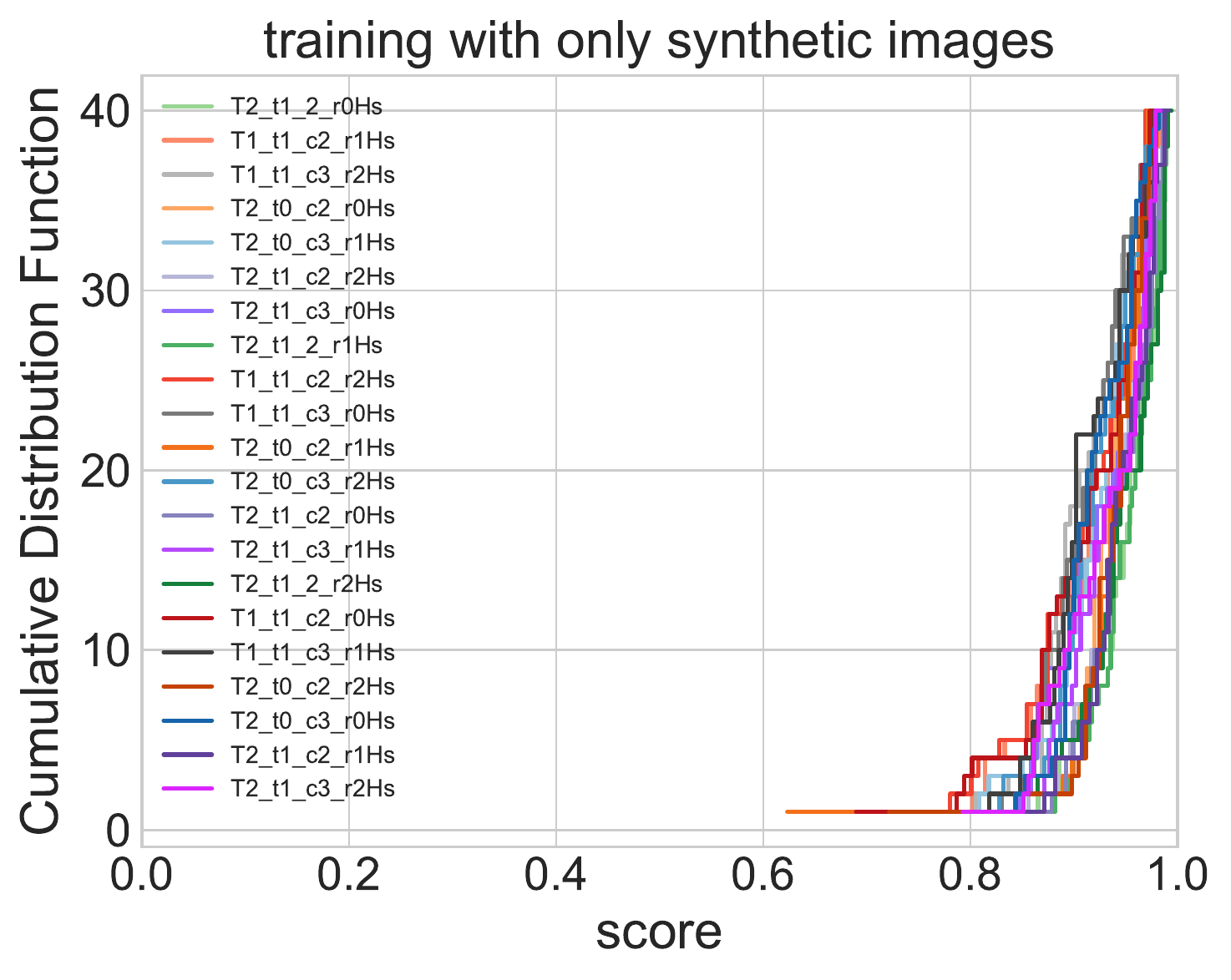}
\includegraphics[width=.90\linewidth]{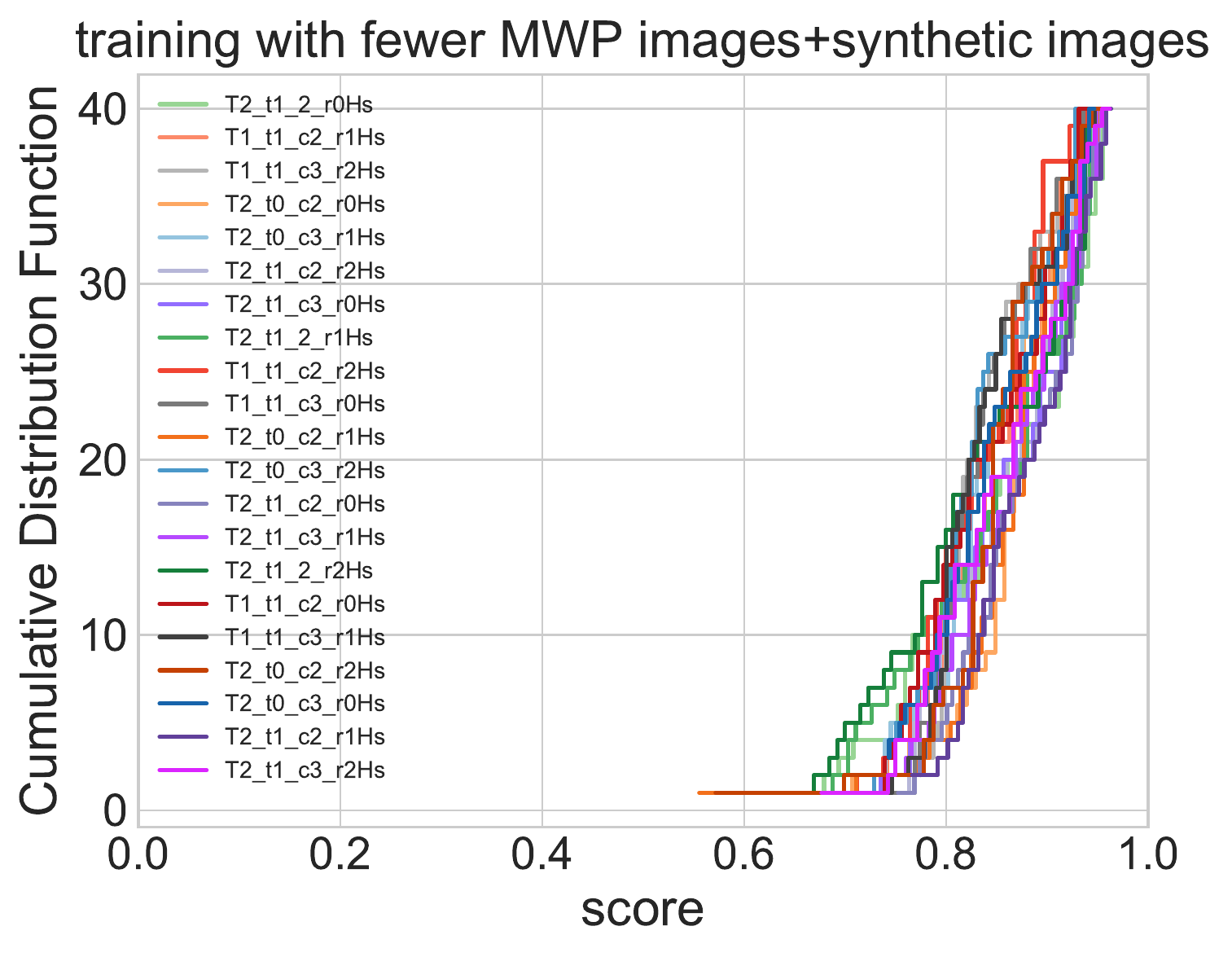}
\caption{The CDF of all the scores given by the retrained algorithm without original MWP training set but with only synthetic images (top panel) and the algorithm retrained on half original MWP training set and the synthetic images (bottom panel). The labels are described in Table~\ref{parameters of rf}. }
\label{fig.socre-cdf-sim-MWP}
\end{figure}

{We next explore the impact of the training set size and composition on the performance of retrained algorithm. We retrain the algorithm with only synthetic images and retrain the algorithm with a set containing half the number of MWP bubbles and all the synthetic images.
Figure~\ref{fig.socre-cdf-sim-MWP} shows the performance of the algorithm trained with only synthetic images and the algorithm trained with fewer MWP images+synthetic images on the noisy data. Compared with the scores returned when training with all the MWP data and synthetic images in Figure~\ref{fig.socre-cdf-noise}, the scores returned by different random forests are similar but more concentrated. This is likely caused by the larger fraction of synthetic bubbles, which are similar to the test set, in the training set. The synthetic images 
are responsible for the better performance of the retrained algorithm on the synthetic images test set. }

The increased scores after retraining suggest the original training dataset is incomplete, especially lacking bubbles driven by intermediate or low-mass stars. We further examine the performance of retrained \Brut\ on observational data in Section~\ref{Re-Testing Brut on Milky Way Project Data} and \ref{Application: Bubbles in the Perseus Molecular Cloud}.

\begin{table*}[]
\begin{center}
\caption{Parameters of the Synthetic Images \label{bubble-parameter}}
\begin{tabular}{ccccccc}
\hline

ID & Label$^{a}$ & Turbulence$^{b}$ & Evolutionary Stage (Myr) & Image Size (pc) & Crop  & Noise \\
\hline
1 &  T1\_t1\_c2          &   T1    &   0.1    &    2.2   &  \cmark &  \xmark \\
2 &  T1\_t1\_c3          &   T1    &   0.1    &     3  &  \cmark  &  \xmark \\
3 &  T2\_t1\_2          &   T2    &   0.1    &    2.2   & \xmark  &  \xmark \\
4 &  T2\_t0\_c2          &    T2   &   0.05    &    2.2   &  \cmark  &  \xmark \\
5 &  T2\_t0\_c3          &    T2    &    0.05   &    3   &  \cmark  &  \xmark \\
6 &  T2\_t1\_c2          &     T2   &  0.1     &    2.2   &  \cmark  &  \xmark \\
7 &  T2\_t1\_c3          &     T2   &  0.1     &     3  &  \cmark   &  \xmark \\
8 &  T1\_t1\_c2n          &   T1    &   0.1    &    2.2   &  \cmark &  \cmark \\
9 &  T1\_t1\_c3n          &   T1    &   0.1    &     3  &  \cmark  &  \cmark \\
10 &  T2\_t1\_2n          &   T2    &   0.1    &    2.2   & \xmark  &  \cmark \\
11 &  T2\_t0\_c2n          &    T2   &   0.05    &    2.2   &  \cmark  &  \cmark \\
12 &  T2\_t0\_c3n          &    T2    &    0.05   &    3   &  \cmark  &  \cmark \\
13 &  T2\_t1\_c2n          &     T2   &  0.1     &    2.2   &  \cmark  &  \cmark \\
14 &  T2\_t1\_c3n          &     T2   &  0.1     &     3  &  \cmark   &  \cmark \\
\hline
\multicolumn{5}{p{0.65\linewidth}}{Notes:}\\
\multicolumn{5}{p{0.65\linewidth}}{ 
$^{a}$ The label with ``n'' indicates the synthetic image with noise.
}\\
\multicolumn{5}{p{0.65\linewidth}}{ $^{b}$ Turbulent distributions listed in Table~\ref{Model Properties}.
}\\

\end{tabular}
\end{center}
\end{table*}

\begin{table*}[]
\begin{center}
\caption{Parameters of the Random Forests \label{parameters of rf}}
\begin{tabular}{cccc}
\hline
Random Forests  &\multicolumn{2}{c}{Training Set} & Test Set \\\cline{2-4}\
 Label $^{a}$ &   MWP Zone$^{b}$ / Number   +& Synthetic Image (ID)$^{c}$ / Number   &  \\
 \hline
T1\_t1\_c2\_r1 & r1 / 314 & no / 0 & T1\_t1\_c2 \\
T1\_t1\_c2\_r1s & r1 / 314 & 1-7 / 280  & T1\_t1\_c2 \\
T1\_t1\_c3\_r1 & r1 / 314 & no / 0& T1\_t1\_c3 \\
T1\_t1\_c3\_r1s & r1 / 314 & 1-7 / 280 & T1\_t1\_c3 \\
T2\_t1\_2\_r1 & r1 / 314 & no / 0& T2\_t1\_2 \\
T2\_t1\_2\_r1s & r1 / 314 & 1-7 / 280 & T2\_t1\_2 \\
T2\_t0\_c2\_r1 & r1 / 314 & no / 0& T2\_t0\_c2 \\
T2\_t0\_c2\_r1s & r1 / 314 & 1-7 / 280 & T2\_t0\_c2 \\
T2\_t0\_c3\_r1 & r1 / 314 & no / 0& T2\_t0\_c3 \\
T2\_t0\_c3\_r1s & r1 / 314 & 1-7 / 280 & T2\_t0\_c3 \\
T2\_t1\_c2\_r1 & r1 / 314 & no / 0& T2\_t1\_c2 \\
T2\_t1\_c2\_r1s & r1 / 314 & 1-7 / 280 & T2\_t1\_c2 \\
T2\_t1\_c3\_r1 & r1 / 314 & no / 0& T2\_t1\_c3 \\
T2\_t1\_c3\_r1s & r1 / 314 & 1-7 / 280 & T2\_t1\_c3 \\
T1\_t1\_c2n\_r1 & r1 / 314 & no / 0& T1\_t1\_c2n \\
T1\_t1\_c2n\_r1s & r1 / 314 & 1-14/ 560 & T1\_t1\_c2n \\
T1\_t1\_c3n\_r1 & r1 / 314 & no / 0& T1\_t1\_c3n \\
T1\_t1\_c3n\_r1s & r1 / 314 & 1-14/ 560 & T1\_t1\_c3n \\
T2\_t1\_2n\_r1 & r1 / 314 & no / 0& T2\_t1\_2n \\
T2\_t1\_2n\_r1s & r1 / 314 & 1-14/ 560 & T2\_t1\_2n \\
T2\_t0\_c2n\_r1 & r1 / 314 & no / 0& T2\_t0\_c2n \\
T2\_t0\_c2n\_r1s & r1 / 314 & 1-14/ 560 & T2\_t0\_c2n \\
T2\_t0\_c3n\_r1 & r1 / 314 & no / 0& T2\_t0\_c3n \\
T2\_t0\_c3n\_r1s & r1 / 314 & 1-14/ 560 & T2\_t0\_c3n \\
T2\_t1\_c2n\_r1 & r1 / 314 & no / 0& T2\_t1\_c2n \\
T2\_t1\_c2n\_r1s & r1 / 314 & 1-14/ 560 & T2\_t1\_c2n \\
T2\_t1\_c3n\_r1 & r1 / 314 & no / 0& T2\_t1\_c3n \\
T2\_t1\_c3n\_r1s & r1 / 314 & 1-14/ 560 & T2\_t1\_c3n \\
T1\_t1\_c2n\_r1Ns & no positive training set$^{d}$ / 0 & 1-14/ 560 & T1\_t1\_c2n \\
T1\_t1\_c2n\_r1Hs & half r1 / 159 & 1-14/ 560 & T1\_t1\_c2n \\
T1\_t1\_c2\_r2 & r2 / 311 & no / 0 & T1\_t1\_c2 \\
T1\_t1\_c2\_r3 & r3 / 311 & no / 0 & T1\_t1\_c2 \\
T1\_t1\_c2n\_r2Hs & half r2 / 161 & 1-14/ 560 & T1\_t1\_c2n \\
T1\_t1\_c2n\_r3Hs & half r3 / 158 & 1-14/ 560 & T1\_t1\_c2n \\
\vdots & \vdots   &  \vdots   & \vdots  \\
\hline
\multicolumn{4}{p{0.65\linewidth}}{Notes:}\\
\multicolumn{4}{p{0.65\linewidth}}{
$^{a}$ We list the random forest ``r1'' and ``r1s'' for example, where suffix ``s'' means adding synthetic images into the training set. The random forests label with only ``s'' adopts the synthetic images without noise as part of the training set. The label with both ``n'' and ``s'' means adding the synthetic images with and without noise into the training set. There is a similar set of cases for random forest ``r2'', ``r2s'', ``r3'' and ``r3s''. {``r1Ns'' indicates the training set only includes the synthetic images without any MWP bubbles in the positive training set. ``r1Hs'' means the training set consists of half the MWP bubble in r1 and all the synthetic images.}   
}\\
\multicolumn{4}{p{0.65\linewidth}}{
$^{b}$ The random forest zone listed in Table~\ref{Random Forest Zone}.
}\\
\multicolumn{4}{p{0.65\linewidth}}{
$^{c}$ The synthetic images listed in Table~\ref{bubble-parameter}. The synthetic images are divided into two equal parts. One half acts as a training data set, and the second half serves as a test set.
}\\
\multicolumn{4}{p{0.65\linewidth}}{
$^{d}$ The training set does not have any MWP bubbles in the positive training set but contains the MWP images without bubbles in the negative training set. 
}\\

\end{tabular}

\end{center}
\end{table*}

\subsection{Re-Testing \Brut\ on the Milky Way Project Data}
\label{Re-Testing Brut on Milky Way Project Data}

\begin{figure}[htp]
\centering
\includegraphics[width=1.0\linewidth]{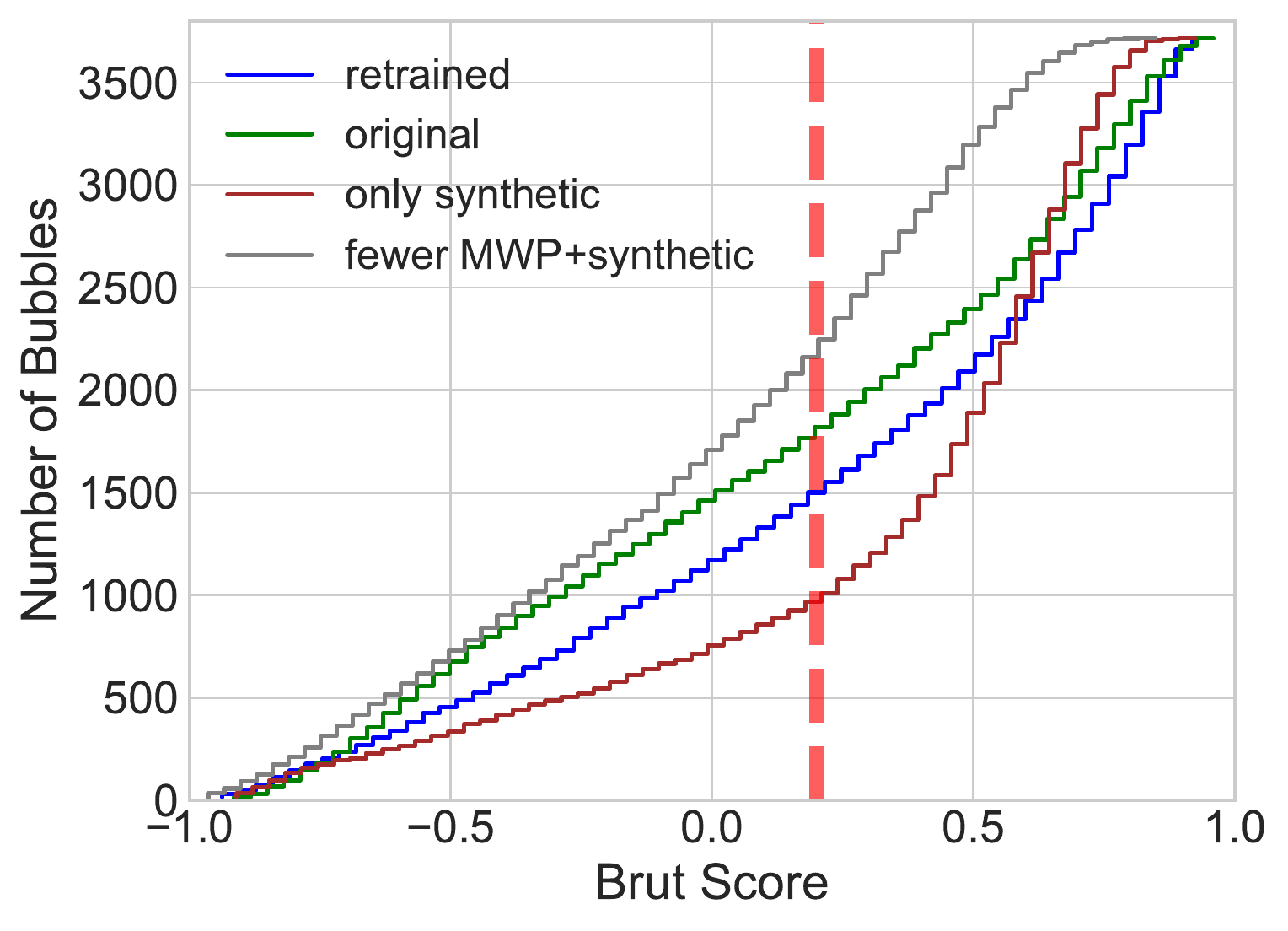}
\caption{The CDF of the scores of 3716 Milky Way Project large bubbles given by algorithm with the original {training} and the algorithm retrained on several different training sets including synthetic bubbles.}
\label{fig.socre-cdf-MWP}
\end{figure}

\begin{figure}[htp]
\centering
\includegraphics[width=1.0\linewidth]{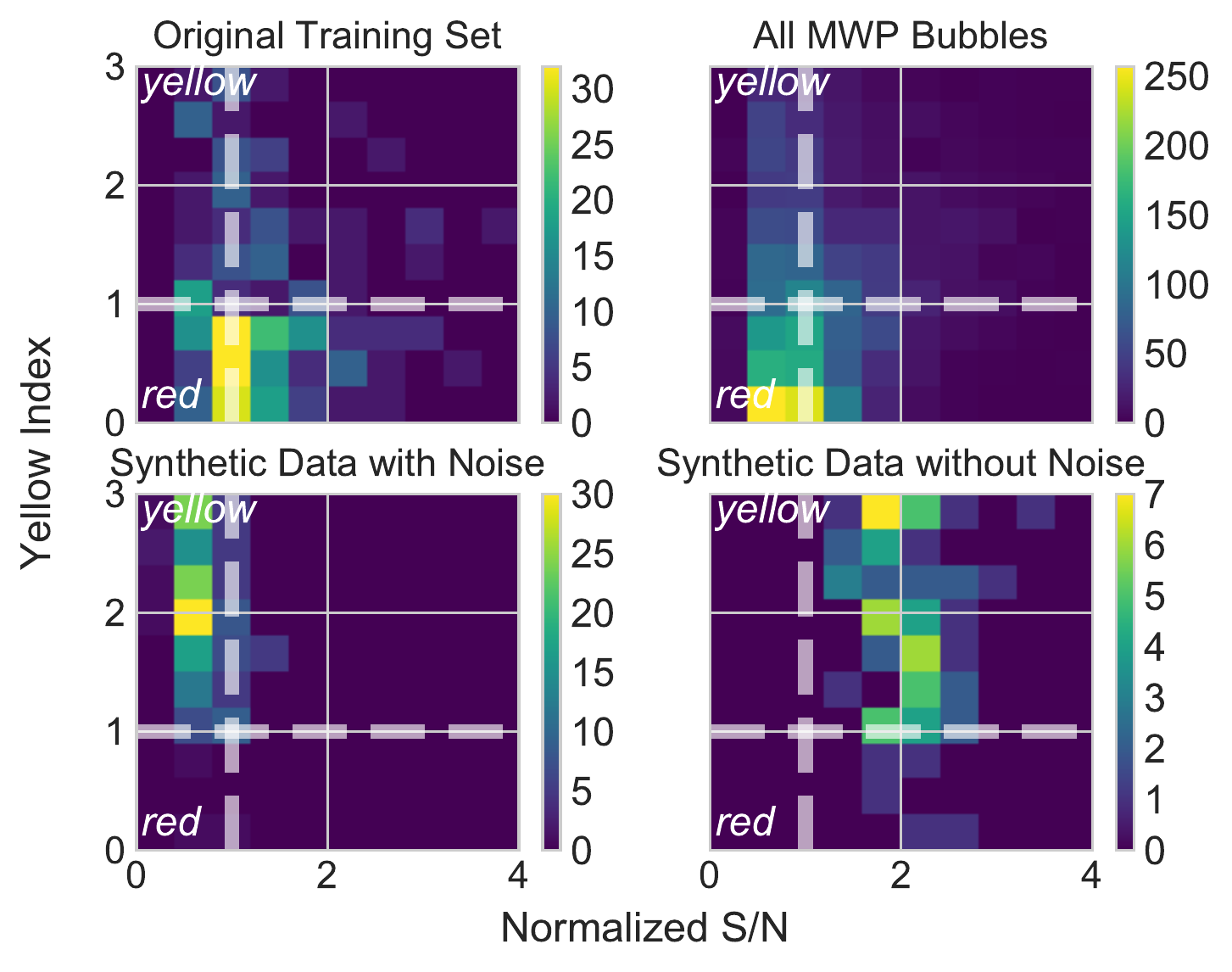}
\caption{Distribution of image properties for four different training sets, where the colors indicate the number of bubbles in each bin. The white dashed lines divides the bubbles into four regions. The upper left quadrant indicates images that contain low S/N yellow bubbles. The upper right quadrant indicates images with high S/N yellow bubbles. The lower left quadrant indicates low S/N red bubbles. The lower right quadrant indicates high S/N red bubbles. }
\label{fig.socre-hist2d}
\end{figure}

We adopt all 3716 large bubbles found by the citizen scientists in \citet{2012MNRAS.424.2442S} as a test set to assess the performance of \Brut\ after retraining. We ignore the objects contained in the ``small bubble" catalogue, which are mainly green knots, dark nebulae, star clusters, galaxies or fuzzy red objects. We compare the performance for the original {training}, the retrained algorithm (using both noisy and noiseless synthetic bubbles){, the algorithm trained with only synthetic images and the algorithm trained with fewer MWP images+synthetic images} in classifying MWP bubbles, as shown in Figure~\ref{fig.socre-cdf-MWP}. The scores returned by the retrained algorithm are significantly higher compared with those returned by \Brut\ without additional training. {When we retrain the algorithm with only synthetic images, the scores under 0.55 show a dramatic improvement. 
After investigating the high and low score bubble images, we find the algorithm trained with only synthetic images improves the scores of ambiguous bubbles with low S/N and reduces the scores of red bubbles with high S/N. 

To explain the performance of the algorithm retrained on several different training sets, we characterize the bubble properties that compose each training set as shown in Figure~\ref{fig.socre-hist2d}. The ``Normalized S/N'' quantifies the contrast and S/N of the image. We define it as 
\begin{equation} \label{eq:Normalized S/N}
{\rm Normalized\,} S/N=C(\bar I_{95}- \bar I_{30})(\bar I_{95}-\bar I_{50})f_{\ge 8  \sigma},
\end{equation}
where $(\bar I_{95}-\bar I_{30})$ is the difference between the top 5\% and the bottom 30\% of values, $(\bar I_{95}-\bar I_{50})$ is the difference between the top 5\% values and the median value, $f_{\ge 8  \sigma}$ is the fraction of bright pixels ($\ge 8 \sigma$), and $C$ is a constant to normalize the values to unity. In most of the high S/N bubble images, the bubble rim structures occupy the top 5\% of the image values, and the noise occupies the bottom 30\%. The average of the diffuse emission is well represented by the median image value. We use this product 
to indicate the contrast of the image. In a random noisy image, 
the normalized S/N is close to 0. The $x$-axis in Figure~\ref{fig.socre-hist2d} indicates the ``Yellow Index,'' which describes the color of the bubble. We define it as the ratio between the number of yellow pixels and the number of red pixels. Although the original training set spans a wide range of color and S/N, it is concentrated in the red domain. The large representation of red bubbles in the training set means that {\it Brut} will more easily identify red bubbles than yellow bubbles. In contrast, the synthetic bubbles are located in the yellow part of the parameter space. The MWP bubbles are mostly low S/N red bubbles, with some low S/N yellow bubbles and high S/N red bubbles, but there are very few high S/N yellow bubbles. Consequently, the algorithm trained with only synthetic images mainly captures bubbles 
with low S/N. This explains why a training set with only synthetic images improves the scores of ambiguous bubbles with low S/N and reduces the scores of bubbles that are red and have high S/N. 

Figure~\ref{fig.socre-cdf-MWP} also shows the result when we randomly remove half of the bubbles in the original MWP training set. The score distribution returned by the algorithm trained with fewer MWP images+synthetic images compared to when the results of the algorithm trained with only synthetic images is surprising.
When including half of the original MWP bubbles in the training set, the performance of the algorithm dramatically decreases. The original MWP training bubbles are mostly red, while the synthetic images nearly all contain yellow bubbles. Consequently, these sets inhabit two different color domains. 
The reduction of red bubbles in the training set lowers the scores of these types of bubbles in the test set.
When including all the original and synthetic images in the training set, the performance of the retrained algorithm significantly and steadily improves. Consequently, this demonstrates the 
composition and size of the training set significantly impacts the performance of the algorithm. 

Following our comparison of the algorithm performance after retraining with several different training sets, 
we adopt the training that includes all the original MWP bubbles and synthetic images. This training set significantly 
improves the scores 
of most bubbles with little change in the number of high-score objects.  Although the algorithm trained with only synthetic images improves the scores of a large number of bubbles, it no longer returns any high score bubbles, which were previously assigned to images with red bubbles. }

The MWP characterizes the consensus among users that an image contains a bubble in terms of the ``hit rate'', which is the fraction of citizen scientists who identified a bubble in the image. They define hit rates above 0.1 as being high-confidence bubble candidates.

\begin{figure}[htp]
\centering
\includegraphics[width=0.950\linewidth]{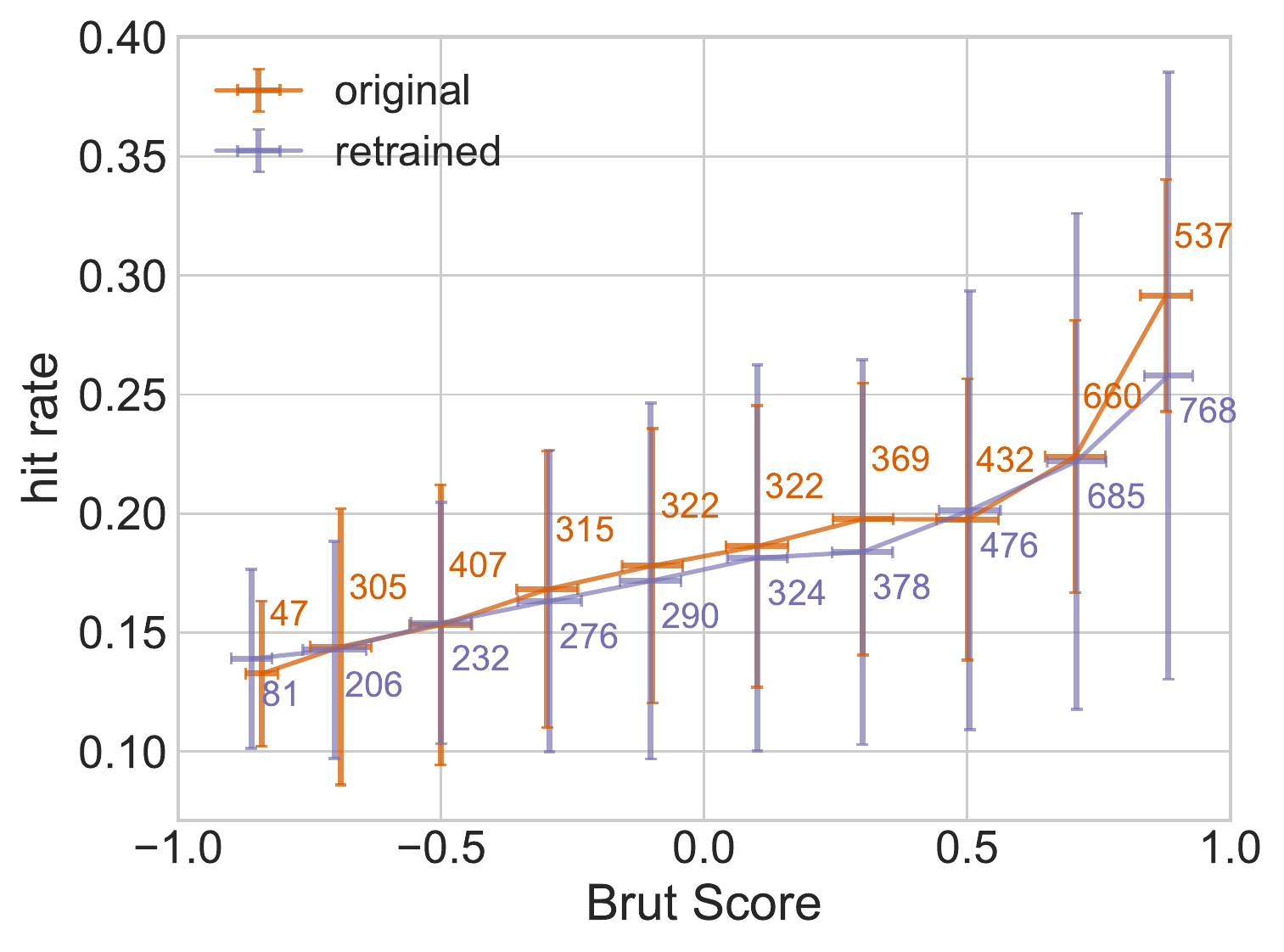} 
\includegraphics[width=0.95\linewidth]{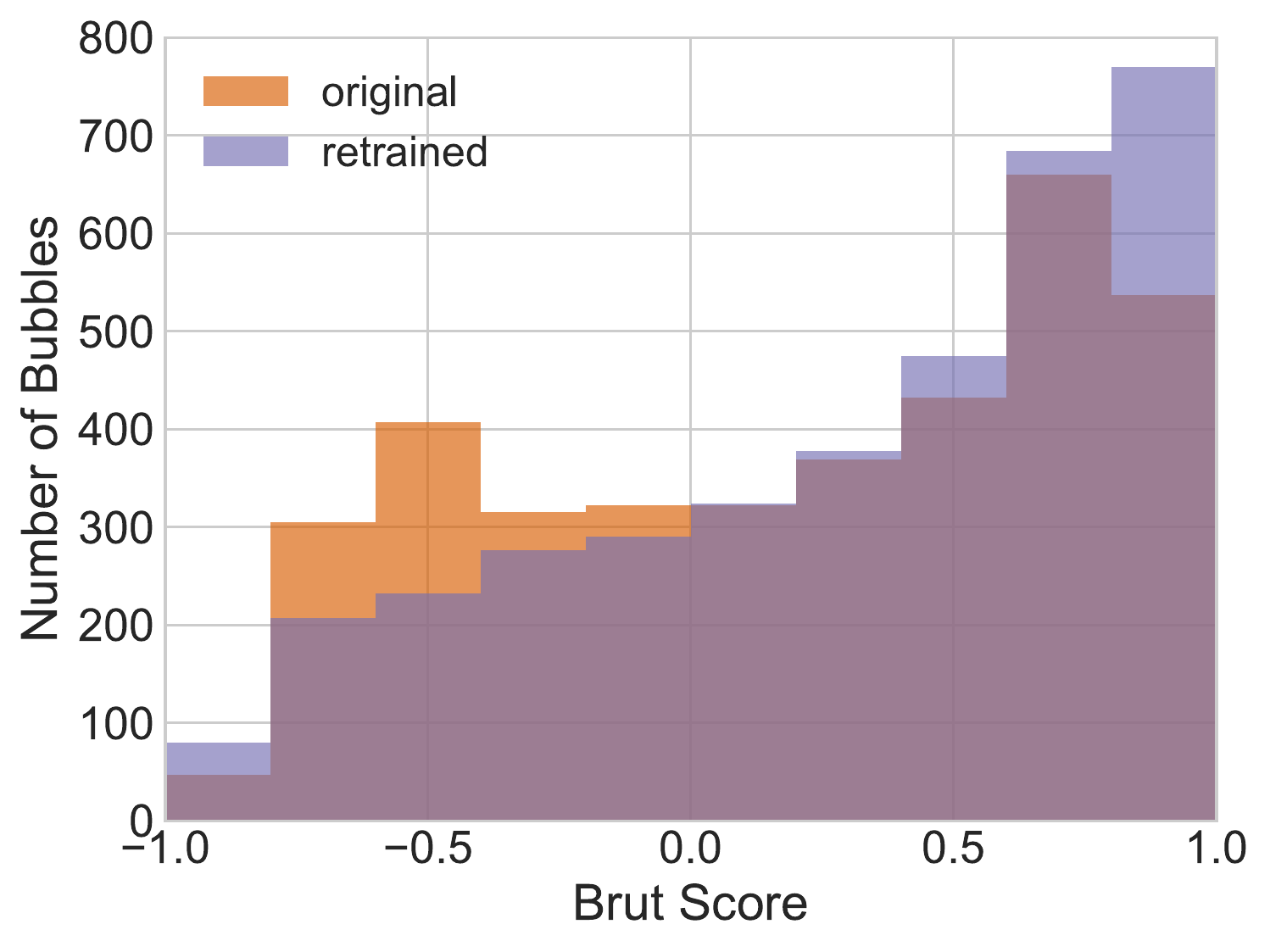} 
\caption{Upper: The average hit rate versus the average binned \Brut\ score for the 3716 MWP large bubbles. The red and green lines indicate the average \Brut\ score returned with the original {training}  and the algorithm retrained on synthetic images both with and without noise, respectively. {The error bars indicate the standard deviation of the scores and hit rate in each bin.} The label indicates the number of bubbles in each bin.  Lower: The distribution of bubble scores returned with the original {training} and after \Brut\ is retrained.}
\label{fig.socre-MPW-hitrate}
\end{figure}

We further compare the scores given by \Brut\ with the original {training}, the scores after it is retrained, and the MWP hit rate as shown in Figure~\ref{fig.socre-MPW-hitrate}. The average \Brut\ score in each bin with the original {training} and after retraining both show a clear trend with the hit rate. {The error bars indicate the standard deviation of the scores and hit rate in each bin.} The higher the hit rate, the higher the score \Brut\ returns, which is consistent with our expectations. In other words, the retrained algorithm preserves the hit-rate distribution, where bubbles with low hit rates continue to have low scores.

Moreover, over 10\% of the MWP bubbles, which were previously marginal or ambiguous detections, are reclassified as high-confidence bubbles after retraining.
Their average \Brut\ score increases from -0.07 to 0.39. About 2\% of the previously identified MWP bubbles are no longer classified as high-confidence bubbles, and their average \Brut\ score drops from 0.31 to 0.06.

\begin{figure*}[htp]
\centering
\includegraphics[width=.99\linewidth]{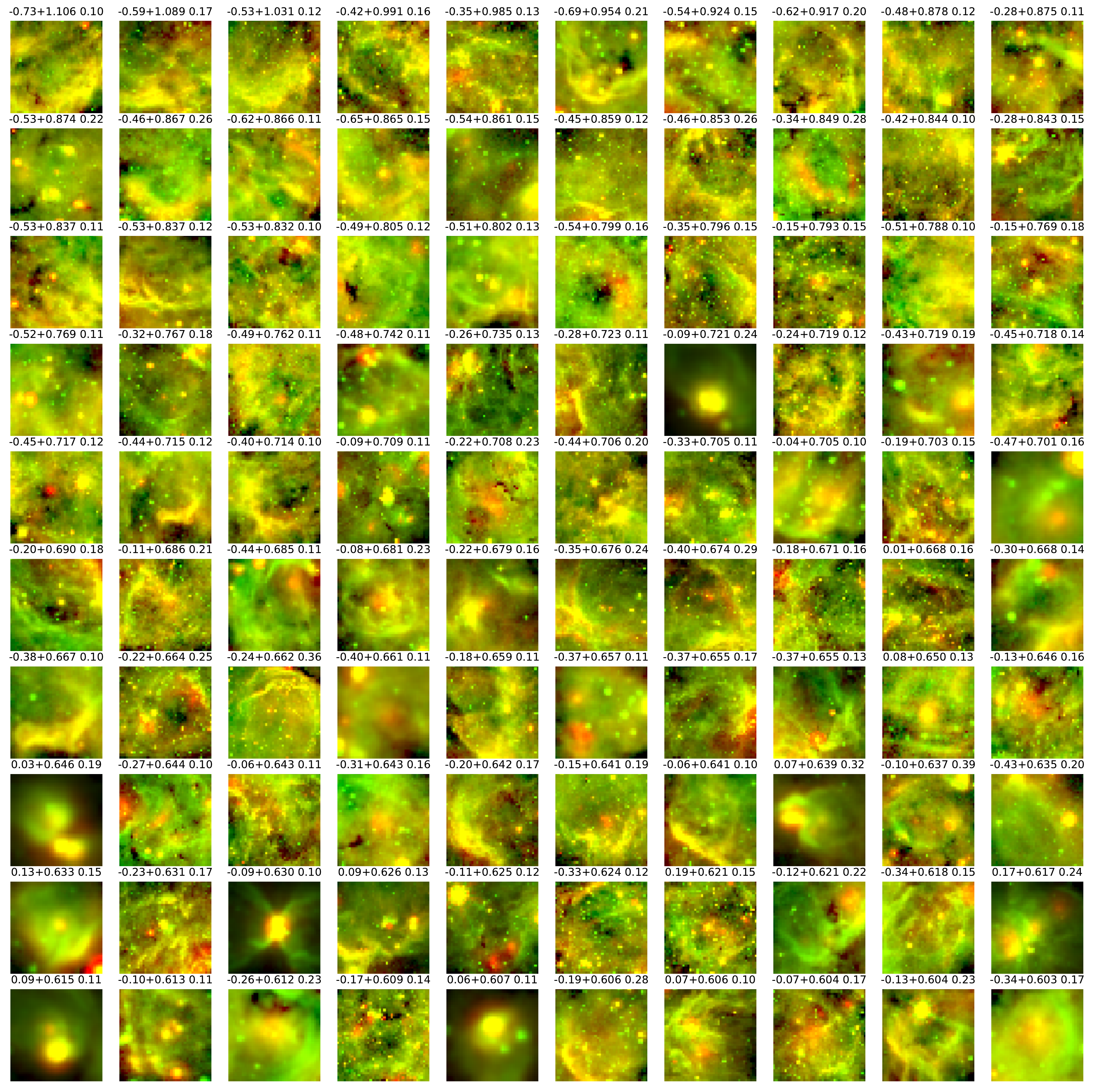}
\caption{One hundred bubbles from MWP. The first number in the title of each panel presents the raw score, which is returned by the original MWP training algorithm. The middle number is the change in score after \Brut\ is retrained with synthetic observations. The last number is the hit rate. }
\label{fig.MPW-new-bubble}
\end{figure*}

Figure~\ref{fig.MPW-new-bubble} shows one hundred bubbles, whose score significantly increases after retraining. Most of these bubbles are yellow, indicating the 8 \um\ and 24 \um\ emission are similar. These yellow bubbles are likely ultra-compact and compact \hii\ regions or analogous regions for less massive B-type stars {\citep{2015ApJ...799..153K}}. The performance of the retrained algorithm is consistent with our training set, in which bubbles are created by the stellar winds of B-type stars. For these type of stars, the amount of ionizing radiation is small, so the bubbles are predominantly cleared by the wind (or earlier protostellar outflows) and then illuminated by the stellar radiation field. 

\begin{figure}[htp]
\centering
\includegraphics[width=.99\linewidth]{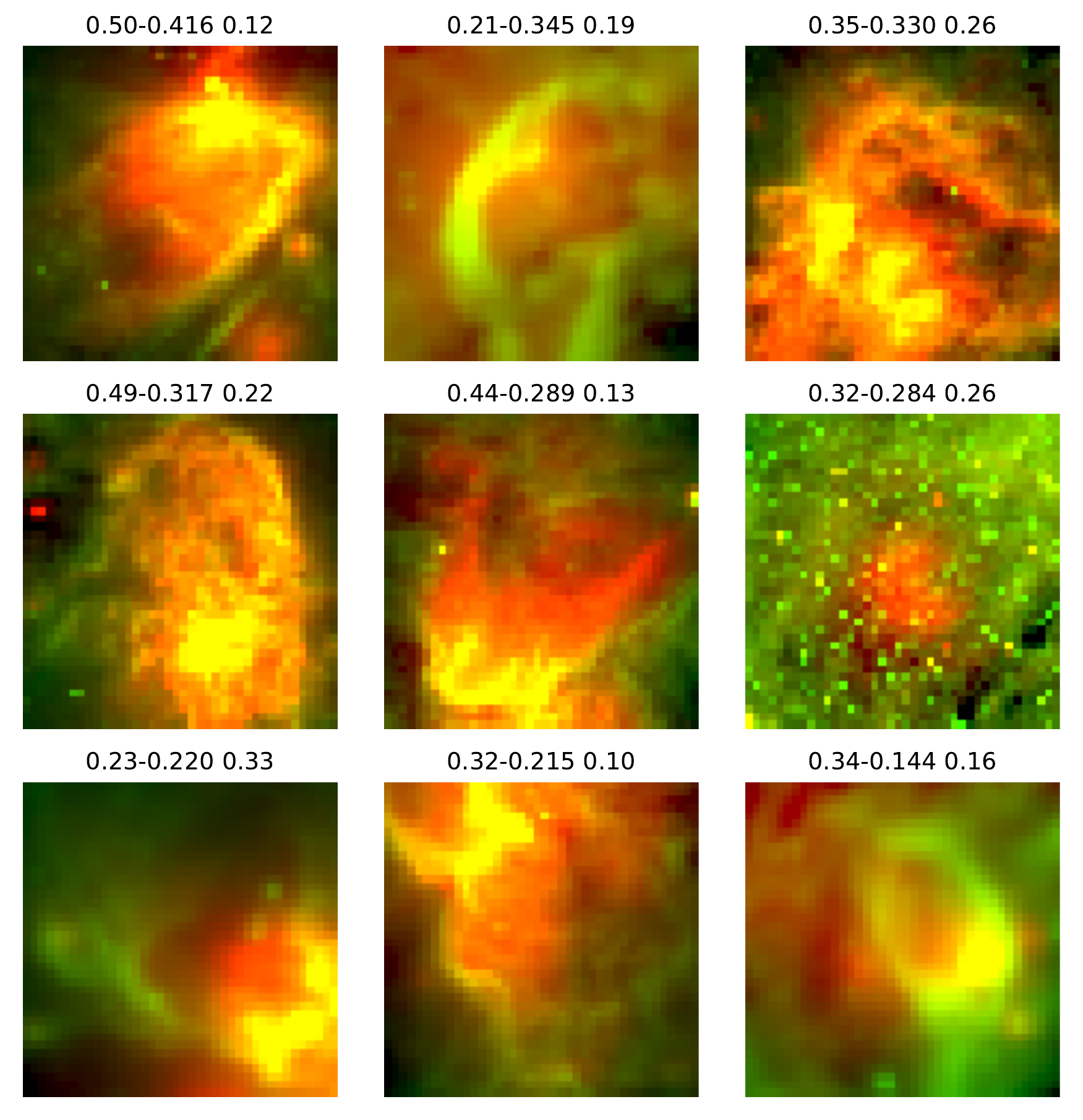}
\caption{Nine bubbles, which were previously likely MWP bubbles but are no longer classified as high-confidence bubbles after retraining. The meaning of each number in the title is described in Figure~\ref{fig.MPW-new-bubble}.  }
\label{fig.MPW-new-bubble-decrease}
\end{figure}

Figure~\ref{fig.MPW-new-bubble-decrease} shows nine bubbles, which were previously identified MWP bubbles but are no longer classified as high-confidence bubbles after retraining. These bubbles 
are very red and, thus, {quite} distinct from our
yellow bubbles, and their morphology does not show a distinct shell rim. Consequently, since we supplemented the training set with synthetic yellow bubbles, the decline of these bubbles \Brut\ scores is unsurprising.

In summary, the performance of the retrained algorithm in classifying yellow bubbles significantly increases when synthetic observations are added to the training set.

\subsection{Application: Bubbles in the Perseus Molecular Cloud}
\label{Application: Bubbles in the Perseus Molecular Cloud}

Perseus is located in the larger Taurus-Auriga-Perseus dark cloud complex with a distance of 250$\pm$50 pc, spanning a total area of about 70 pc$^{2}$ \citep{2006ApJ...638..293E,2009ApJS..181..321E}. With a mass of 10$^{4}$ \msun, the Perseus cloud is often considered to be an intermediate case between low-mass star forming regions such as Taurus and turbulent, high-mass regions such as Orion \citep{1994ApJ...433..117L}, making it an ideal location to study low and intermediate-mass star formation. The feedback of young stars makes Perseus a ``bubbly" cloud \citep{2011ApJ...742..105A}.

\begin{figure}[htp]
\centering
\includegraphics[width=.49\linewidth]{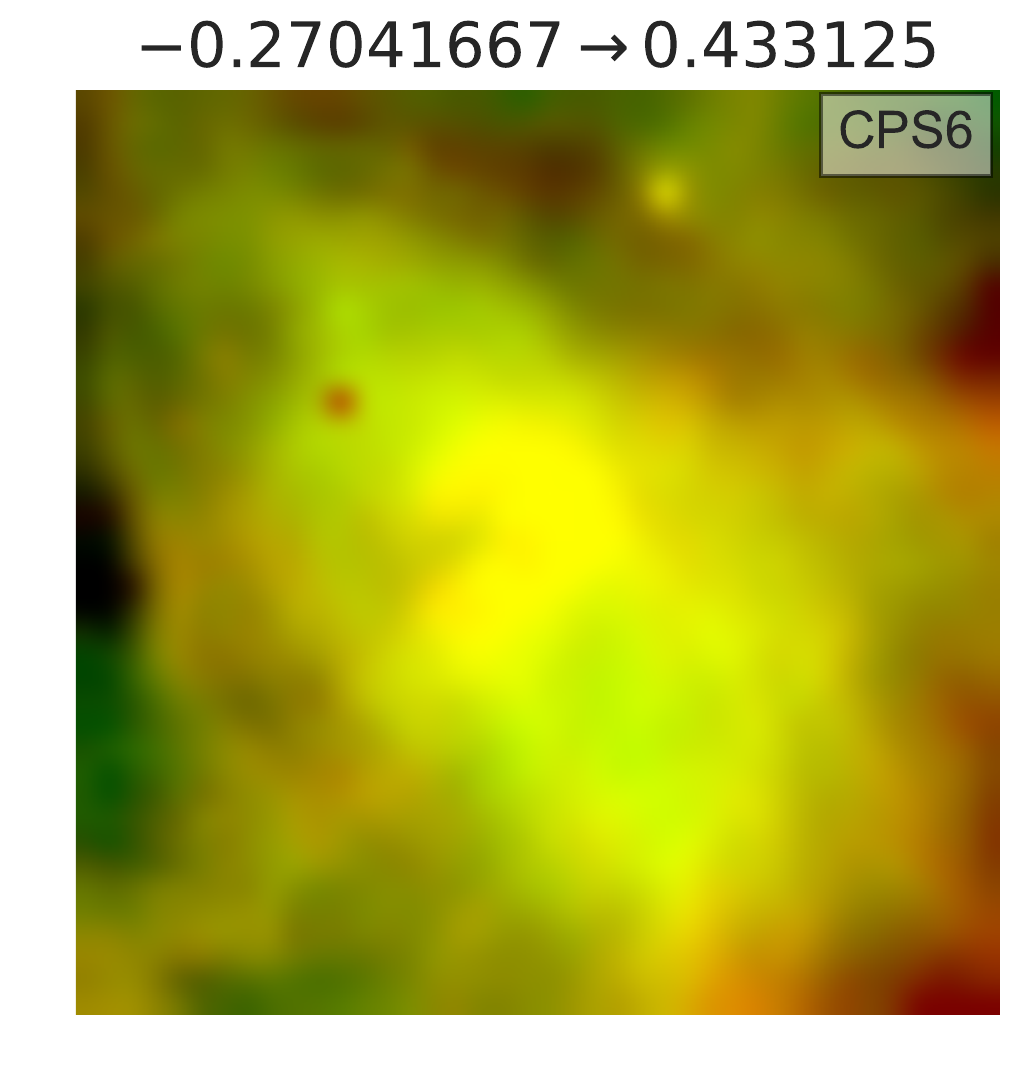}
\includegraphics[width=.49\linewidth]{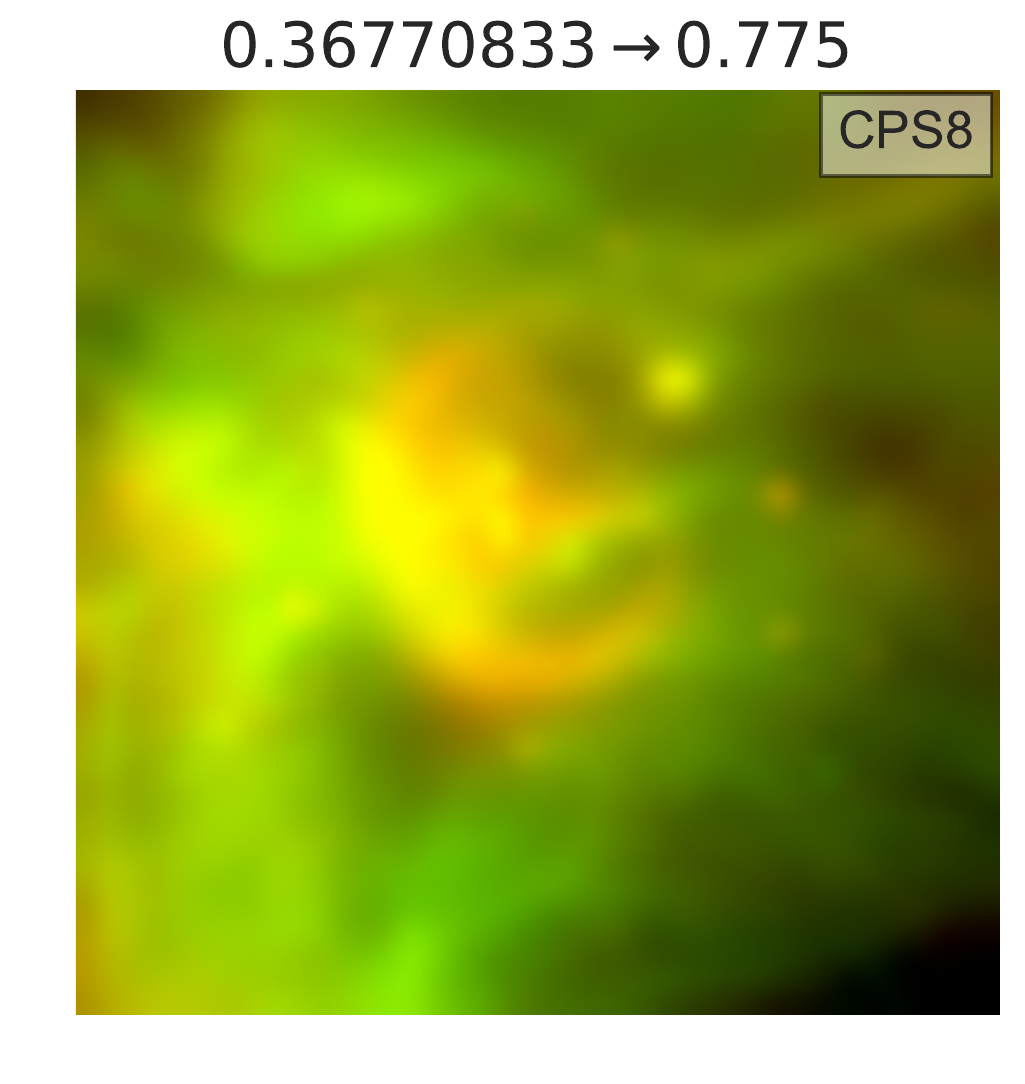}
\includegraphics[width=.49\linewidth]{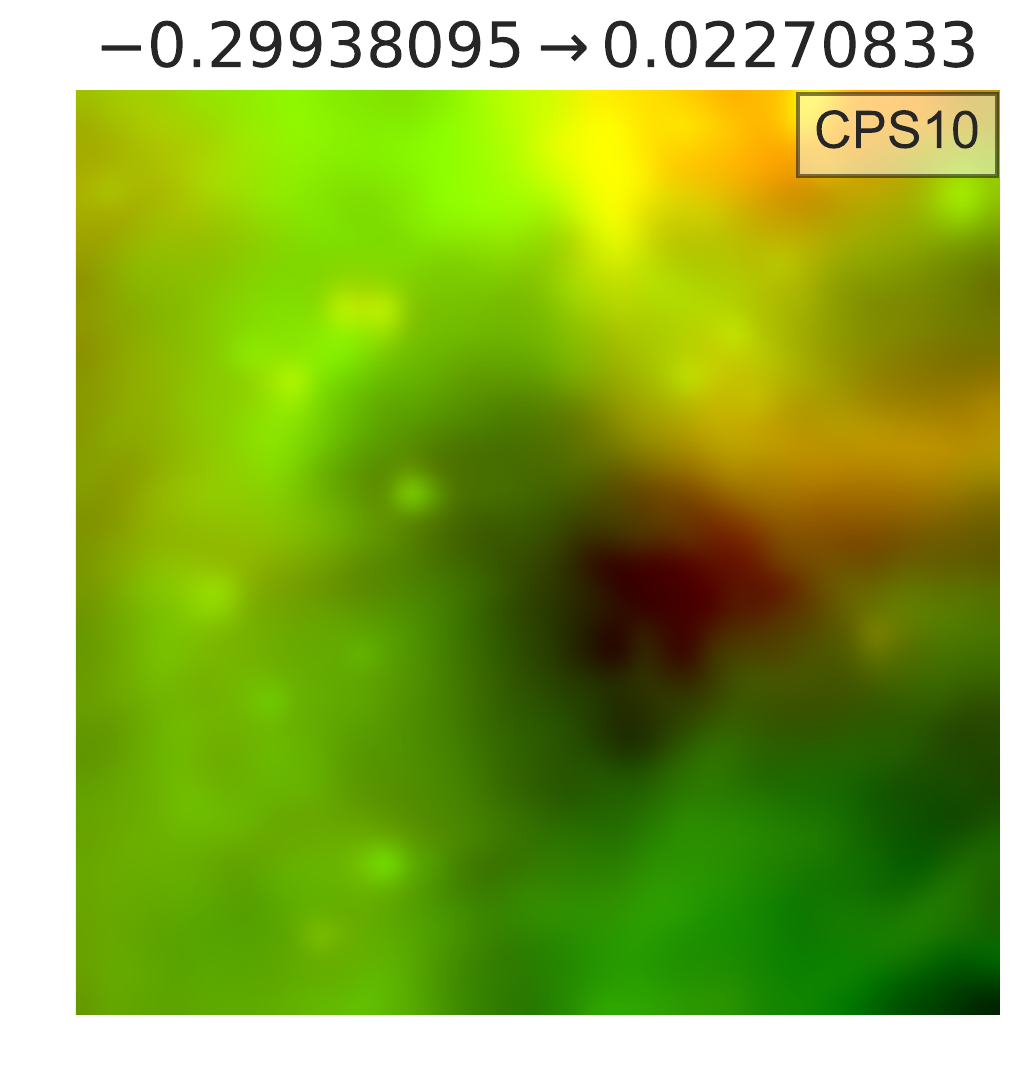}
\includegraphics[width=.49\linewidth]{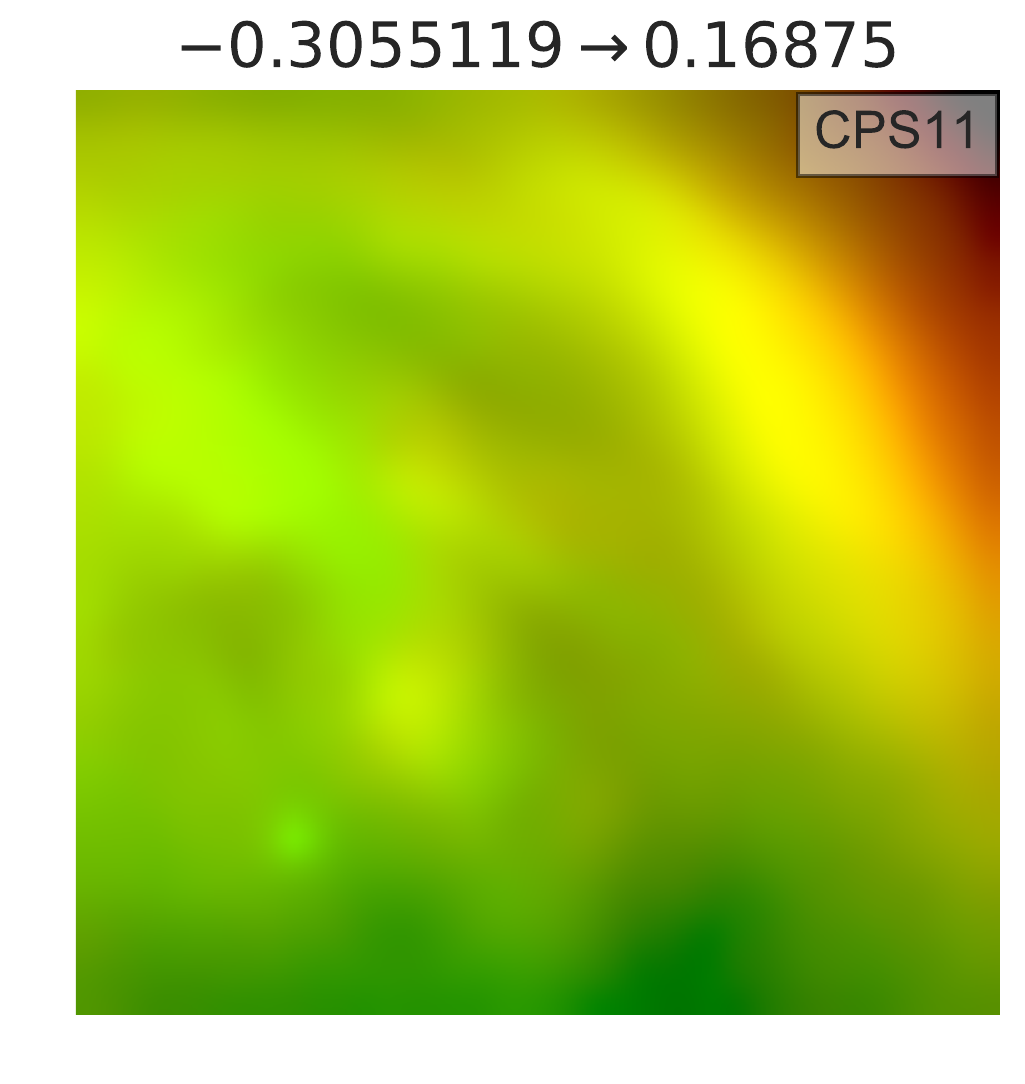}
\caption{Four examples of bubbles in the Perseus molecular cloud. The upper right label in each panel corresponds to the bubble name in \citet{2011ApJ...742..105A}. The left number in the title of each panel indicates the raw score, which is returned by the original {training} algorithm. The right number in the title of each panel indicates the new score returned by the retrained algorithm.}
\label{fig.perseus-bubble-cps}
\end{figure}

\citet{2011ApJ...742..105A} identified 12 bubbles using CO spectral data. We extract the Spitzer image of Perseus in 4.5 \um, 8 \um\ and 24 \um\ bands (Gutermuth priv.~comm.) and apply \Brut\ to this data. Figure~\ref{fig.perseus-bubble-cps} shows four examples of bubbles in the Perseus molecular cloud. These bubbles are associated with shells CPS6, CPS8, CPS10 and CPS11 in the CO data, which were visually identified by \citet{2011ApJ...742..105A}. Table~\ref{perseus-bubble-parameter} lists the physical properties of these bubbles. All these bubbles are probably driven by relatively low or intermediate-mass young stars such as B type or F type stars. Figure~\ref{fig.perseus-bubble-cps} shows these bubbles and their associated \Brut\ scores before and after retraining. These four cases show a significant improvement in score, most from a negative score (non-bubble) to a positive score (likely bubble).

CPS6 and 8 are similar to the synthetic bubbles and the MWP yellow bubbles. They are the best examples of the good performance produced by retraining \Brut. CPS11 is a partial bubble, which is probably why its score is still $<$ 0.2. Table~\ref{perseus-bubble-parameter} shows that CPS10 is driven by a B5V star, but there is no distinct evidence of the existence of the star in the infrared images. However, in the optical data, the star is bright and is clearly visible. 

The dust emission exceeds that of the star, so the B5V star becomes invisible when embedded in the cloud. Although CPS10 is not a yellow bubble, it is nonetheless consistent with the bubble model in Figure 1 in \citet{2014ApJS..214....3B}, where green shell structure is produced by PAH emission and the red interior is dominated by hot dust. The bubble score is low, which is likely due to the contamination by other emission at the upper right corner.

These results indicate the retrained algorithm can perform well for molecular cloud data not included in the MWP. The synthetic observations are able to improve \Brut\ performance in classifying bubbles produced by relatively low or intermediate-mass young stars such as B-type stars.

\begin{table*}[]
\begin{center}
\caption{Physical Properties of Four Perseus Bubbles \label{perseus-bubble-parameter}}
\begin{tabular}{ccccc}
\hline

Bubble & Cloud& Center & Candidate & Source   \\
Name & Region &($\alpha_{2000},\delta_{2000}$) & Source & Type  \\
\hline
CPS-6         &   L1468    &   03 41 24, 31 54 10    &    IRAS 03382+3145   & unknown$^{a}$ \\
CPS-8         &  IC 348    &   03 44 10, 32 17 20    &    omi Per   & B1III $^{b}$ \\
CPS-10          &   IC 348    &   03 44 35, 32 10 10    &    HD 281159   &  B5V $^{c}$\\
CPS-11         &   IC 348      &   03 44 50, 32 18 10    &     V* 695 Per \& IC 348 LRL 30   &  M3.75 \& F0 $^{d}$ \\
\hline
\multicolumn{5}{p{0.65\linewidth}}{Notes:}\\
\multicolumn{5}{p{0.65\linewidth}}{
$^{a}$ This is likely not a typical main-sequence star; it is probably a young pre-main-sequence star \citep{2011ApJ...742..105A}. There is no reliable mid-IR detection in the MIPS bands because the source is confused with the bright nebulosity in this region \citep{2007ApJS..171..447R}.
}\\
\multicolumn{5}{p{0.65\linewidth}}{
$^{b}$ This bright source is observed both in the optical and the infrared, and it is classified as a YSO candidate in the c2d point-source catalog \citep{2009ApJS..181..321E}.
}\\
\multicolumn{5}{p{0.65\linewidth}}{
$^{c}$. HD 281159 is a binary system with two massive B5 main sequence stars, which has a disk around the binary pair. And they have an age $\le$ 10 Myr \citep{mora01}.
}\\
\multicolumn{5}{p{0.65\linewidth}}{
$^{d}$ Two possible candidates are V* 695 Per and IC 348 LRL 30, which are an M3.75 star and F0 star, respectively. Both stars are classified as a YSO candidates in the c2d catalog and both have been detected in X-ray \citep{2001AJ....122..866P}. 
}

\end{tabular}
\end{center}
\end{table*}

\section{Conclusions}
\label{Conclusion}

We adopt magneto-hydrodynamics simulations of stellar winds interacting with a molecular cloud and post-process them using a three-dimensional dust continuum Monte-Carlo radiative transfer code. We generate synthetic observations of bubbles in the Spitzer bands (4.5 \um, 8 \um\ and 24 \um). We employ a previously developed machine learning algorithm, \Brut\, and quantitatively evaluate its performance in identifying bubbles using synthetic dust observations. Our main findings are the following:

\begin{enumerate}[1.]
\item Synthetic observations in combination with visually identified sources can be used to significantly improve machine learning classification.

\item After {retraining} with synthetic images, \Brut\ better identifies yellow bubbles, which are likely associated with \hii\ regions for less massive B-type stars or cavities evacuated by stellar winds.

\item { The completeness of the training set significantly impacts the performance of the algorithm.} We suggest that the number of yellow bubbles in the current MWP bubble catalog is incomplete, and we expect a random search of the full {GLIMPSE} dataset with \Brut\ would return many more yellow bubble candidates.

\item Some of the bubbles with improved scores are associated with lower confidence sources in the MWP. These would likely be identified as bubbles by an expert, and thus the simulations provide an efficient means to enhance machine learning training sets.

\item Turbulent structures greatly affect the morphology of bubbles, yielding a variety of bubble shapes. Different evolutionary stages and different cropped image sizes further enhance the bubbles contained 
in the training set. Adding noise similar to that in the {GLIMPSE} data makes the synthetic observations more realistic. In combination, these modifications create a more complete training set to improve the machine learning classifications.

\item The retrained algorithm performs well classifying bubbles associated with more embedded sources located in Perseus. Thus, retraining with synthetic observations expands the parameter space of the training set beyond the less embedded and more distant regions with massive stars covered by the MWP.

\end{enumerate}

\acknowledgments
\begin{acknowledgements}
We thank Rob Gutermuth and Chris Beaumont for helpful comments and contributions. {We also thank the anonymous referee for comments that improved this manuscript.} SO acknowledges support from NSF AAG AST-1510021 and NASA ATP NNX15AT05G.
\end{acknowledgements}

\end{CJK*}

\end{document}